\documentclass{aa}  
\usepackage{graphicx}  
\usepackage{txfonts}  
\def\msol{\hbox{\kern 0.20em $M_\odot$}}  
\def\lsol{\hbox{\kern 0.20em $L_\odot$}}  
\def\rsol{\hbox{\kern 0.20em $R_\odot$}}  
\def\sr{\hbox{\kern 0.20em sr}}  
\def\srmu{\hbox{\kern 0.20em sr$^{-1}$}}  
   
\def\g{\hbox{\kern 0.20em g}}  
\def\gmu{\hbox{\kern 0.20em g$^{-1}$}}  
\def\kg{\hbox{\kern 0.20em kg}}  
\def\pc{\hbox{\kern 0.20em pc}}  
   
\def\mum{\hbox{\kern 0.20em $\mu$m}}  
\def\mumd{\hbox{\kern 0.20em $\mu$m$^{-2}$}}  
\def\cm{\hbox{\kern 0.20em cm}}  
\def\m{\hbox{\kern 0.20em m}}  
\def\km{\hbox{\kern 0.20em km}}  
\def\nm{\hbox{\kern 0.20em nm}}  
   
\def\s{\hbox{\kern 0.20em s}}  
\def\h{\hbox{\kern 0.20em h}}  
\def\sec{\hbox{\kern 0.20em sec}}  
\def\min{\hbox {\kern 0.20em min}}  
\def\smu{\hbox{\kern 0.20em s$^{-1}$}}  
\def\smd{\hbox{\kern 0.20em s$^{-2}$}}  
\def\an{\hbox{\kern 0.20em an}}  
\def\anmu{\hbox{\kern 0.20em an$^{-1}$}}  
\def\deg{\hbox{\kern 0.20em $^{\rm o}$}}  
\def\yr{\hbox{\kern 0.20em yr}}  
\def\yrmu{\hbox{\kern 0.20em yr$^{-1}$}}  
\def\Myr{\hbox{\kern 0.20em Myr}}  
\def\Mymu{\hbox{\kern 0.20em Myr$^{-1}$}}  
\def\K{\hbox{\kern 0.20em K}}  
\def\pcmu{\hbox{\kern 0.20em pc$^{-1}$}}  
\def\pcmd{\hbox{\kern 0.20em pc$^{-2}$}}  
\def\pcmt{\hbox{\kern 0.20em pc$^{-3}$}}  
\def\kms{\hbox{\kern 0.20em km\kern 0.20em s$^{-1}$}}  
\def\kmpd{\hbox{\kern 0.20em km$^{2}$}}  
\def\kpc{\hbox{\kern 0.20em kpc}}  
\def\cms{\hbox{\kern 0.20em cm\kern 0.20em s$^{-1}$}}  
\def\erg{\hbox{\kern 0.20em erg}}  
\def\ergs{\hbox{\kern 0.20em erg}}  
\def\cmpd{\hbox{\kern 0.20em cm$^2$}}  
\def\cmmd{\hbox{\kern 0.20em cm$^{-2}$}}  
\def\cmms{\hbox{\kern 0.20em cm$^{-6}$}}  
\def\cmpt{\hbox{\kern 0.20em cm$^3$}}  
\def\cmmt{\hbox{\kern 0.20em cm$^{-3}$}}  
\def\mpd{\hbox{\kern 0.20em m$^2$}}  
\def\mmd{\hbox{\kern 0.20em m$^{-2}$}}  
\def\mpt{\hbox{\kern 0.20em m$^3$}}  
\def\mmt{\hbox{\kern 0.20em m$^{-3}$}}  
\def\mujy{\hbox{\kern 0.20em $\mu$Jy}}  
\def\mjy{\hbox{\kern 0.20em mJy}}  
\def\Mj{\hbox{\kern 0.20em MJy}}  
\def\jy{\hbox{\kern 0.20em Jy}}  
\def\ghz{\hbox{\kern 0.20em GHz}}  
\def\srmd{\hbox{\kern 0.20em sr$^{-1}$}}  
\def \kms{km~$\rm{s}^{-1}$}  
  
\def \mum{$\mu$m}

\begin{document}  
\title{Interstellar Extinction and Long-Period Variables in the \hspace{2cm} Galactic Center}  
% \subtitle{Interstellar extinction and Long-Period Variables}  

\author{M. Schultheis\inst{1,2}  
      \and  
      K. Sellgren\inst{3}  
      \and  
       S. Ram\'{\i}rez\inst{4} 
        \and 
       S. Stolovy\inst{5} 
       \and  
      S. Ganesh\inst{6}  
       \and 
       I.S. Glass \inst{7}
       \and
       L. Girardi\inst{8}  
       }

\offprints{M. Schultheis}

\institute{Observatoire de Besan\c{c}on, 41bis, avenue de l'Observatoire, F-25000 Besan\c{c}on, France 
              \and 
              Institut d'Astrophysique de Paris, CNRS, 98bis Bd Arago, F-75014 Paris, France\\ 
              \email{mathias@obs-besancon.fr} 
         \and  
	  Astronomy Department, Ohio State University, Columbus, OH 43210, USA\\  
	  \email{sellgren@astronomy.ohio-state.edu}  
         \and  
             IPAC, Caltech, Pasadena, CA 91125, USA \\  
             \email{solange@ipac.caltech.edu}  
         \and  
             Spitzer Science Center, Caltech, Pasadena, CA 91125, USA \\ 
             \email{stolovy@ipac.caltech.edu}  
	  \and  
	  Physical Research Laboratory, Astronomy \& Astrophysics Division, Ahmedabad, India\\  
	  \email{shashi@prl.res.in}
          \and 
          South African Astronomical Observatory, PO Box 9, Observatory 7935, South Africa 
	  \and  
	  Osservatorio di Padua, Italy 
	     }  
  
             %\thanks{ }}  
%	  \and  
%	  The OHIO State University, Columbus, OH 43210  
%	  \email{sellgren@astronomy.ohio-state.edu}  
%	  \and  
%	  Physical Research Laboratory, Astronomy \&Astrophysics Division, Ahmedabad, India\\  
%	  \email{shashi@prl.res.in}  
	    
%	  \and  
%	  Osservatorio di Padua\\  

        \date{received:??; accepted ?? }

    \titlerunning{Interstellar Extinction and Long-Period Variables in the Galactic Center} 
    \authorrunning{Schultheis et al.} 
  
% \abstract{}{}{}{}{}  
% 5 {} token are mandatory  
  
\abstract  
{} {To derive a new map of the interstellar extinction near the Galactic
Center (GC) extending to much higher values of $A_V$ than previously
available. To use the results obtained to better characterise the long-period
variable star population of the region.}
{We take the Spitzer IRAC catalogue of GC point sources (Ram\'{\i}rez et al.
2008) and combine it with new isochrones (Marigo et al. 2008) to derive
extinctions based on photometry of red giants and asymptotic giant branch
(AGB) stars.  We apply it to deredden the LPVs found by Glass et al. (2001)
(Glass-LPVs) near the GC. We make period-magnitude diagrams and compare them
to those from other regions of different metallicity.}
 {Our new extinction map of the GC region covers 2.0\deg\ $\times$ 1.4\deg\
(280 $\times$ 200 pc at a distance of 8 kpc). The Glass-LPVs follow
well-defined period-luminosity relations (PL) in the IRAC filter
bands at 3.6, 4.5, 5.8, and 8.0 $\mu$m.  The period-luminosity
relations are similar to those in the Large Magellanic Cloud, suggesting that the 
PL relation in the IRAC bands is universal.  We use ISOGAL data to derive
mass-loss rates and find for the Glass-LPV sample some 
correlation between mass-loss and pulsation period, as expected theoretically.  Theoretical isochrones
for a grid of different metallicities and ages are able to reproduce this
relation. The GC has an excess of high luminosity and long period LPVs
compared to the Bulge, which supports previous suggestions that it contains
a younger stellar population.
}  
   {} 
 
  \maketitle

\section{Introduction}  
  
The variable stars  of the Galactic Center (GC) region are of great
interest, both as population and distance indicators. Long-period,
large-amplitude variables situated on the asymptotic giant branch (AGB),
comprising Miras and OH/IR stars, are the easiest objects to detect thanks
to their high luminosities. Thus, they are one of the few stellar
populations that can be observed in their entirety towards the GC.  As is
well-known, in the innermost parts of the Galaxy, surveys are hampered by high interstellar extinction (e.g. Frogel et al.
\cite{Frogel99}; Schultheis et al. \cite{Schultheis99}); observations must
therefore be carried out in the infrared where $A_{\lambda}$ can be as low
as 0.04 $A_V$ for $4\mu$m $<\lambda<$ $8\mu$m (Indebetouw et al 2005).
Sensitive surveys at the latter wavelengths have only recently become
possible thanks to the Spitzer satellite.

For these reasons, Glass et al. (\cite{Glass2001}) conducted a $K$-band (2.2
$\mu$m) survey for variable stars covering 24 $\times$ 24\,arcmin$^{2}$ (56
$\times$ 56 pc at a distance of 8 kpc) and centered on the GC in a study
spanning 4 years. The majority of the variable sources they found were, as
expected, Miras and OH/IR stars with periods ranging from 150\,d to about
800\,d. Uncertainty as to the foreground extinction unfortunately precluded
any detailed comparison of their luminosities with similar populations in
other well-studied areas, such as the the solar neighbourhood, the Baade's
Windows and the Magellanic Clouds, where period-luminosity relations have
been determined.

The inner Galaxy has also been searched  intensively in the radio
region for OH sources (see e.g. Lindqvist et al. \cite{Lindqvist92},
Sjouwerman et al. \cite{Sjouwerman98}, Wood et al. \cite{Wood98},
Blommaert et al. \cite{Blommaert98}, Vanhollebeke et al.
\cite{vanhollebeke06}). From these surveys, it is clear that a number
of the large amplitude variables remained undetected in the Near-infrared, owing
to the extremely high extinction in some regions. Nevertheless, from OH and
SiO observations, (see e.g. Messineo \cite{Messineo04} and Deguchi et al. \cite{Deguchi08}) radial velocities are available for many of these sources, making them extremely valuable for studies of stellar kinematics near the GC.  

Because the large-amplitude variables have very high mass-loss rates they
also play an important role in the chemical evolution of the Galaxy. It is
therefore desirable that they should be characterised as fully as possible.

Previous studies in the near-IR using 2MASS or DENIS data have yielded
extinction maps with a typical resolution of several arcmin (see e.g. 
Schultheis et al. \cite{Schultheis99}, Marshall et al. \cite{Marshall2006}). 
These maps, however, were limited by the sensitivities of the near-IR surveys
and are only realistic in regions where $A_{V}$ is less than 25 magnitudes.
Unfortunately, this limit is generally exceeded towards the GC.  
  
In this paper we show that an improved map of the interstellar extinction
around the GC can be made by fitting mid-IR data on AGB/RGB stars obtained
by the IRAC camera on Spitzer to isochrones (see also Ganesh et al., in
preparation).  Extinctions have been determined for much more heavily
obscured areas than was possible previously, especially towards the GC
itself. We apply the new extinction values to a discussion of the
long-period variables in the region, concentrating on the period-magnitude
relations that they obey in the IRAC bands and their mass-loss rates.

\section{The data set}  
  
\subsection{The Spitzer IRAC Point Source Catalog of the Galactic Center
(GALCEN)}
  
The central $\rm 2.0\deg\times 1.4\deg$ of the GC have been mapped with
Spitzer/IRAC between 3.6 and 8.0 $\mu$m (Stolovy et al. 2006 and Stolovy et
al. 2008, in preparation). Ram\'{\i}rez et al. (2008) performed point-source
extraction on the IRAC data and published a confusion-limited catalogue of
point sources that also included photometry from 2MASS. The IRAC magnitudes
are referred to as [3.6], [4.5], [5.8] and [8.0], corresponding to their
central wavelengths. The average confusion limits are 12.4, 12.1, 11.7 and
11.2 magnitudes, respectively, but can vary by 2 or 3 magnitudes within the
survey. The whole catalogue (referred to here as GALCEN, see Ram\'{\i}rez et
al. 2008 for more details) consists in total of about one million sources.
Most of these show the characteristics of red giants or AGB stars, but there
are several hundreds of extremely red sources which may be massive Young
Stellar Objects (YSOs). A major problem that must be faced in the study of
the LPV component is the bright-star limit of the IRAC camera. According to
the Spitzer Observer's Manual (Version 7.1), saturation starts at 7.9, 7.4,
4.8 and 4.8 magnitudes respectively. Saturated sources are however retained
in the GALCEN catalogue but if a flux is greater than the limit it is
flagged with the number `3' (see Ram\'{\i}rez et al. 2008 for a more
detailed discussion).  We have excluded photometry with flux flag values of 3
from our analysis.

\subsection{GALCEN vs GLIMPSE-II} 
 
Recently, the GALCEN IRAC observations have been processed independently by
the GLIMPSE-II team (see http://www.astro.wisc.edu/sirtf/docs.html). 
Different data reduction and source extraction procedures from those of the
GALCEN survey were used (see http://www.astro.wisc.edu/glimpse and Benjamin et al. \cite{Benjamin03}). These data were kindly made available to us in
advance of publication by the GLIMPSE-II team.  As discussed in the online
document, they found systematic offsets between the GALCEN and GLIMPSE-II
photometry that was larger than the combined uncertainty in both
observations, at the bright and faint ends of the observed range.  To look
for any possible systematic effects on our work we have also calculated the extinction
maps that we discuss in Sect. 3 using the GLIMPSE-II results.  We find
that the differences in the values of the extinctions derived from GALCEN and
from GLIMPSE-II are smaller than the uncertainties in the determinations.
 
The locations of the LPVs in the IRAC colour-colour diagrams, the IRAC
colour-magnitude diagrams, and the period-luminosity relations (see Sections
4--5) also differ slightly between the GALCEN and GLIMPSE-II photometry. 
The GALCEN and GLIMPSE-II photometry differ by $\sim$0.05 mag in the mean,
with a standard deviation of 0.12 mag, over the brightness range of the
LPVs.  We find no statistically significant dependence of the difference
between the GALCEN and GLIMPSE-II photometry on magnitude, or on the periods
of the LPVs (which can be regarded as a proxy for their de-reddened
magnitudes). The scatter in this difference, however, is large and we
conservatively place a limit of 0.2 mag on any possible variation in its
value over the brightness range occupied by the LPVs. Our conclusions in
this paper are, however, unaffected by photometric uncertainties at this
level. 
 
\subsection{Long-Period Variables}  
  
The Glass et al. (\cite{Glass2001}) survey covered an area of 24 $\times$ 24
arcmin$^{2}$ around the GC, as mentioned, and found about 400 periodic
variables with Mira-like amplitudes and an average period of 427 d. (Among
these are 64 OH/IR stars included in the Wood et al. 1998 sample).  We refer
to the survey list hereafter as the Glass-LPV catalogue. Compared to
the well-explored SgrI Baade's window of low extinction which lies at $l =
-1.4^\circ$ and $b = -2.6^\circ$, the GC field contains more than ten
times the number of variables per arcmin$^2$. The average period in the SgrI
field is much lower, at 333 d. Due to the limited photometric precision of
the Glass et al. (2001) survey, small amplitude variables such as semi-regulars were not
detected.
  
\subsection{ISOGAL observations}  
  
The details of the ISOGAL observational procedure with ISOCAM (Cesarsky et
al. \cite{Cesarsky96}) and the general processing of the data are described
in Schuller et al. (\cite{Schuller2003}). In this paper we will discuss only
the ISOGAL FC$-00027-00006$ field (hereafter called FC$-027$), centered at
$(l,b)=(-0.27^{\circ},-0.06^{\circ})$. It was observed with the 
narrow filters LW5 (6.5--7.1\,$\mu$m) and LW9 (13.9--15.9 \,$\mu$m), using
3\arcsec\ $\times$ 3\arcsec\ pixels, in order to avoid saturation. More specific
details about this field can be found in Schuller et al.
(\cite{Schuller2006}). The parts of the region closest to the GC itself
could not be observed by ISOGAL due to saturation problems even  when
the narrow-band filters were used. Nevertheless, 25\% of the Glass-LPV area
is overlapped by the ISOGAL survey at 7 and 15 $\mu$m, mainly towards the
south-western corner.
  
%\subsection{Near-Infrared spectra}  
  
%Schultheis et al. (\cite{Schultheis2003}) conducted a near-IR spectroscopic
%follow-up program of a sample of 120 sources drawn from this gion with
%mid-IR excesses at 7 and 15\,$\mu$m. Using bolometric luminosities and CO
%and water absorption band data, they were able to identify AGB stars and red
%giant stars as well as young stellar objects (YSOs) and supergiants.  They
%derived metallicities and gravities for their sample using the equivalent
%widths of Na I and Ca I.

\begin{figure*}[ht!]  
\centering  
\includegraphics[angle=0,height=10cm]{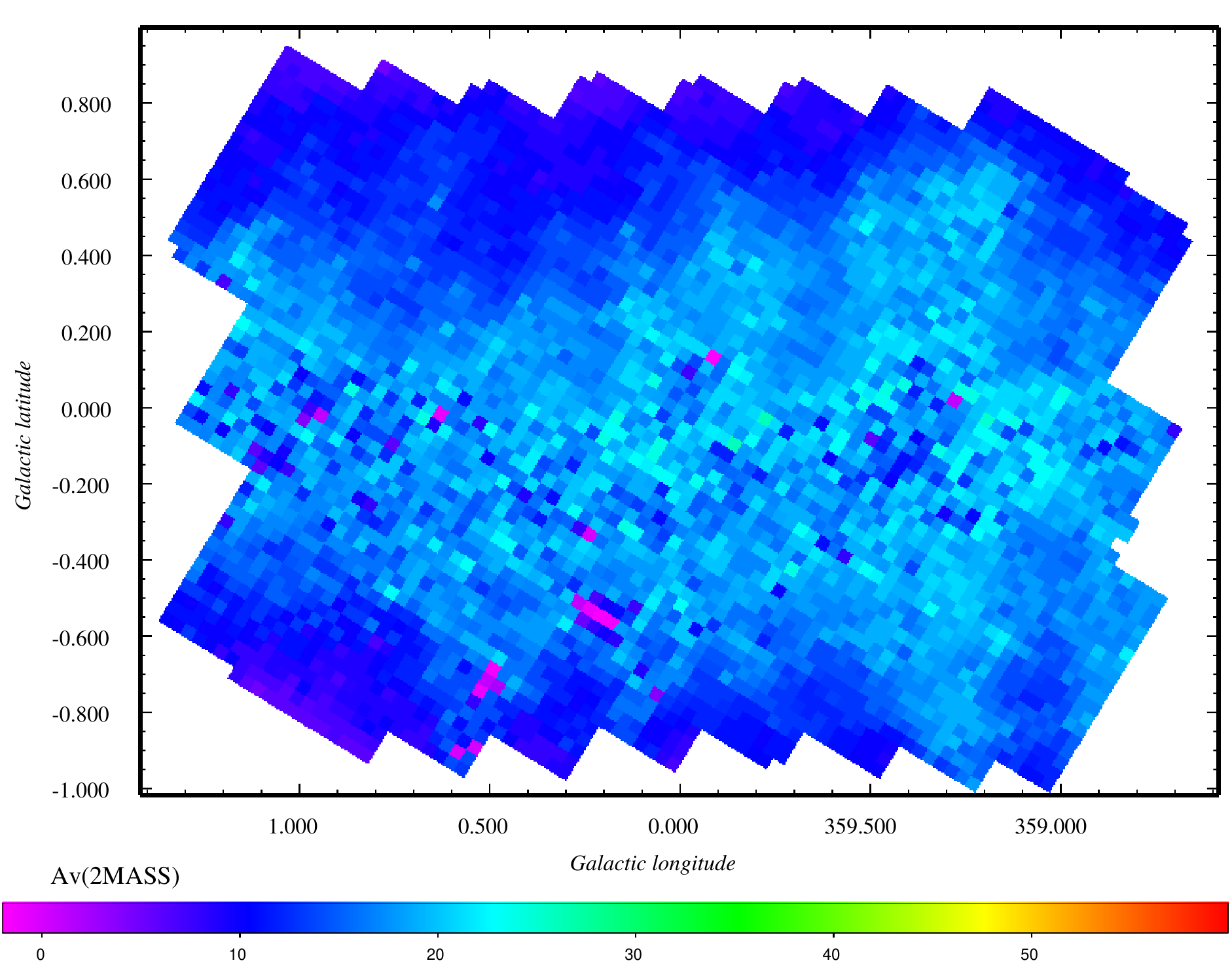}  
\includegraphics[angle=0,height=10cm]{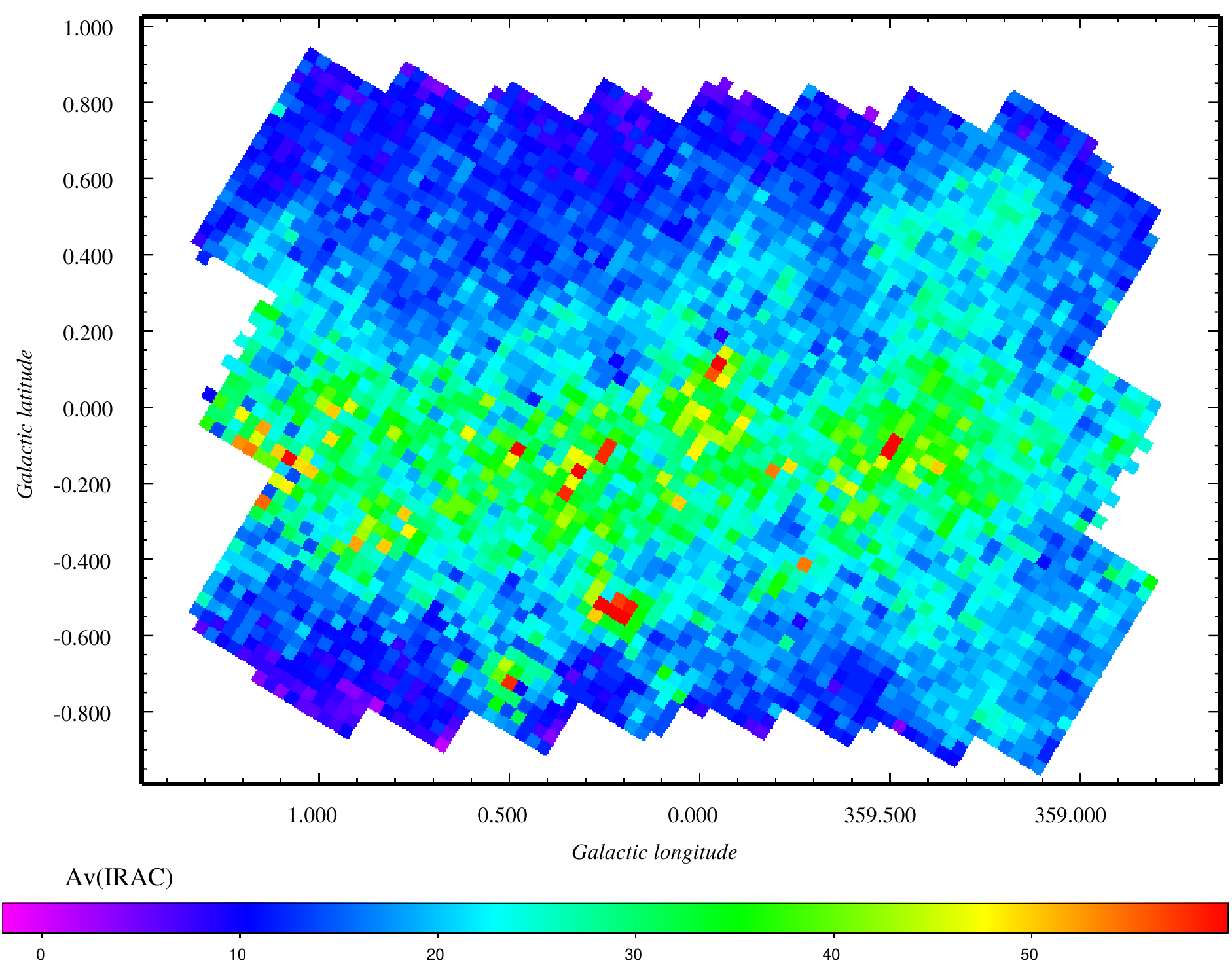}  
\caption{{\em {Upper panel:}} Interstellar extinction map of the central  
280 $\times$ 200\,pc of the Galactic Center with spatial resolution 
2\arcmin, derived using 2MASS $J-K_s$ data.  {\em{Lower panel:}} As above, 
except using the [3.6]$-$[5.8] colour-magnitude data for point sources 
from the Ram\'{\i}rez et al. (2008) GC catalogue.} 
%For $A_{V} < 20$ the 2MASS map is more reliable while for higher extinction 
%regions the IRAC map should be used} 
 
\label{ext2MASS}  
\end{figure*}

\subsection{Cross-identifications}  
  
Cross-identification of stellar catalogues in these extremely crowded regions,
which are often confusion limited, is a very difficult issue. In order to
minimize spurious associations we used the following method when comparing
the different catalogues mentioned above. We first calculated the offsets of
the 400 brightest sources between a given pair of catalogues and determined
the second-order distortion matrix. Following application of the offset and
distortion corrections, we made the subsequent cross-identifications using a
search radius of 2 arcsec:

\begin{itemize} 
  
\item First, we cross-identified the LPVs of Glass et al. (2001) with the
GALCEN catalogue. By limiting the GALCEN catalogue to the area covered by
the Glass-LPV survey, it was reduced to a total of 46314 sources. Of the 421
LPVs, 410 were found in the GALCEN list (97\%).  Many of them are, however,
saturated, especially at [3.6] and [4.5].  (From the Glass-LPV sample there were 162,
178, 36, and 58 saturated sources at [3.6], [4.5], [5.8], and [8.0],
respectively).
 
\item Second,  
we cross-identified the Glass-LPV catalogue with ISOGAL. We found 183 matches,
despite the small overlap.
  
%\item Third, we cross-identified  the near-IR spectroscopic sample of 
%Schultheis et al. (\cite{Schultheis2003}) with the GALCEN catalogue, resulting 
%in 108 matches.
%The GALCEN fluxes are saturated at [3.6], [4.5], [5.8], and 
%[8.0] for 50, 60, 13, and 20 of these sources, respectively. 
  
\end{itemize}  
 
\section{Interstellar extinction}  
  
\subsection{Derivation of the extinction values}

As stated, interstellar extinction remains a serious obstacle to the
interpretation of stellar populations in the GC region.   

The present work follows the method of Ganesh et al. (in preparation) who
recently mapped the extinction towards the inner Bulge with a spatial
resolution of 2\arcmin. GLIMPSE-II and 2MASS data are compared to the latest
isochrones for evolved stars (Marigo et al. \cite{Marigo2007}) in order to
derive $A_V$. Several studies of this type, using only the 2MASS data, have
been undertaken before (e.g. Schultheis et al. \cite{Schultheis99},
Marshall et al. \cite{Marshall2006}).

In the method of Ganesh et al, colour-magnitude diagrams are constructed
within sampling boxes of 2\arcmin\ square and the amount by which each
individual data point has to be de-reddened for it to fall on the isochrone
is determined. The extinction values for $A_{V} > 20$ are derived using the
[3.6]$-$[5.8] colour, while 2MASS $J$ and $K_s$ data are used for $A_{V} <
20$.

With the large number of filter combinations available (three from 2MASS and
four from GLIMPSE-II) several possible combinations had to be considered.
The [5.8] vs. [3.6]$-$[5.8] colour-magnitude diagram was selected since it
was found to yield the smallest dispersion, i.e. the
$[3.6]-[5.8]$ colour excess was found to be the most suitable one for
determining the interstellar extinction (see e.g. Indebetouw et al.
\cite{Indebetouw05}).  This pair of filters has, in fact, the advantage of
simultaneous observations (the other two IRAC colours were observed at
different times) and also has better completeness (more stars detected
through high extinction). Further, at 3.6\,$\mu$m and 5.8\,$\mu$m, the
observed flux is dominated by stellar photospheric emission while at
8.0\,$\mu$m interstellar PAH emission becomes prominent.  The models
indicate that metallicity effects are negligible in the [3.6]$-$[5.8] colour,
in contrast to $J - K_s$, where the RGB/AGB branch of a metal-rich stellar
population shifts to redder $J - K_s$ colours (Schultheis et al.
\cite{Schultheis2004}). For more detailed information see Ganesh et al. (in
preparation).
  
In the following, we took the GALCEN catalogue of Ram\'{\i}rez et
al. (\cite{Ramirez2008}) and limited ourselves to stars with
$\rm [3.6]$ $< 12.5$ and $\rm [5.8]$ $< 12$.
We have also excluded sources with [3.6]--[4.5] $>$ 0.8 and [5.8]--[8.0] $>$ 0.5
from our extinction calculation, because sources with these colors
are typical of YSOs (see Allen et al. 2004) and the inclusion of their
extreme colors would bias the results. We again chose a sampling box of
2\arcmin\  square in order to get a sufficient numbers of stars for the isochrone
fitting. For each sampling box the average extinction value was taken. 

The values of $A_{\lambda}/A_V$ that we use to make the extinction maps were
derived by using the infrared colour-colour diagram $J-K_{s}$ vs.
$K_{s}-$[IRAC] as proposed by Jiang et al. (\cite{Jiang03}), Indebetouw et
al.  (\cite{Indebetouw05}) and Ganesh et al. (2008). We used the $J-K_{s}$ vs. $K_{s}-$[3.6] and the  $J-K_{s}$ vs. $K_{s}-$[5.8] diagrams to determine
$\rm A_{[3.6]}$ and $\rm A_{[5.8]}$. We assumed that most of the sources are luminous RGB stars or AGB stars with moderate mass-loss and similar intrinsic
 colours. We further adopted the values $A_{J} = 0.256 \times A_{V}$ and $A_{K_{S}} = 0.089 \times A_{V}$ as used by Schultheis et al. (\cite{Schultheis99}). 
 Fitting straight lines in these diagrams gives us the following values: $\rm A_{[3.6]}/A_{V} =0.0498 \pm 0.0015$ and $\rm A_{[5.8]}/A_{V} = 0.0308 \pm 0.0015$. {These values were determined using the whole GALCEN catalog. 
Ganesh et al. (2008) will discuss more in detail how the extinction coefficients vary with different line of sights using the whole GLIMPSE-II dat set.  Our $\rm A_{[3.6]}/A_{V}$ value agrees within the errors with those from
 Indebetouw et al. (2005), Lutz (1999), Flaherty et al. (\cite{Flaherty07}) and Rom{\'a}n-Z{\'u}{\~n}iga et al.
\cite{Roman07}, while our  derived $\rm A_{[5.8]}/A_{V}$ is a litte bit lower, but still lies inside the errors of the quoted authors.

Lutz et al. (1996) and Lutz (1999)  measured a flat mid-IR extinction curve towards the GC. 
However Flaherty et al. (\cite{Flaherty07}) and Rom{\'a}n-Z{\'u}{\~n}iga et al.
\cite{Roman07} have shown that extinction coefficents may vary somewhat
depending in which line of sight one is looking (e.g. especially towards star forming regions). Recently, Chapman et al. (2008) have also studied the mid-infrared extinction law in three molecular clouds and found that as the overall extinction increases, the  curve flattens out. The coefficients evidently vary according to whether the  dust is  located in the general ISM or in dense molecular clouds.

Figure \ref{ext2MASS} shows the $A_{V}$ map derived from the 2MASS filters
(upper panel) and that from the IRAC [3.6] and [5.8] filters of GALCEN
(lower panel).  There is a great similarity in the overall distribution of
the two extinction maps. Where 2MASS gives lower limits for the extinction
or is unable to probe the extinction due to lack of sources, the GALCEN
extinction map exhibits high values for the extinction ($A_{V} > 40$). 
Within Figure ~\ref{ext2MASS} there are regions of low extinction according
to the 2MASS data but which are seen to have very high extinction when
looked at by IRAC (e.g. the region around $l=0.2\deg$ and $b=-0.5\deg$). 
These areas are mostly associated with infrared dark clouds. 2MASS gives
artificially low values of extinction in such cases because it detects only
those stars located in front of the dark cloud whereas IRAC
penetrates to much greater depth.

\begin{figure}[h!] 
\centering  
\includegraphics[angle=0,width=6.5cm]{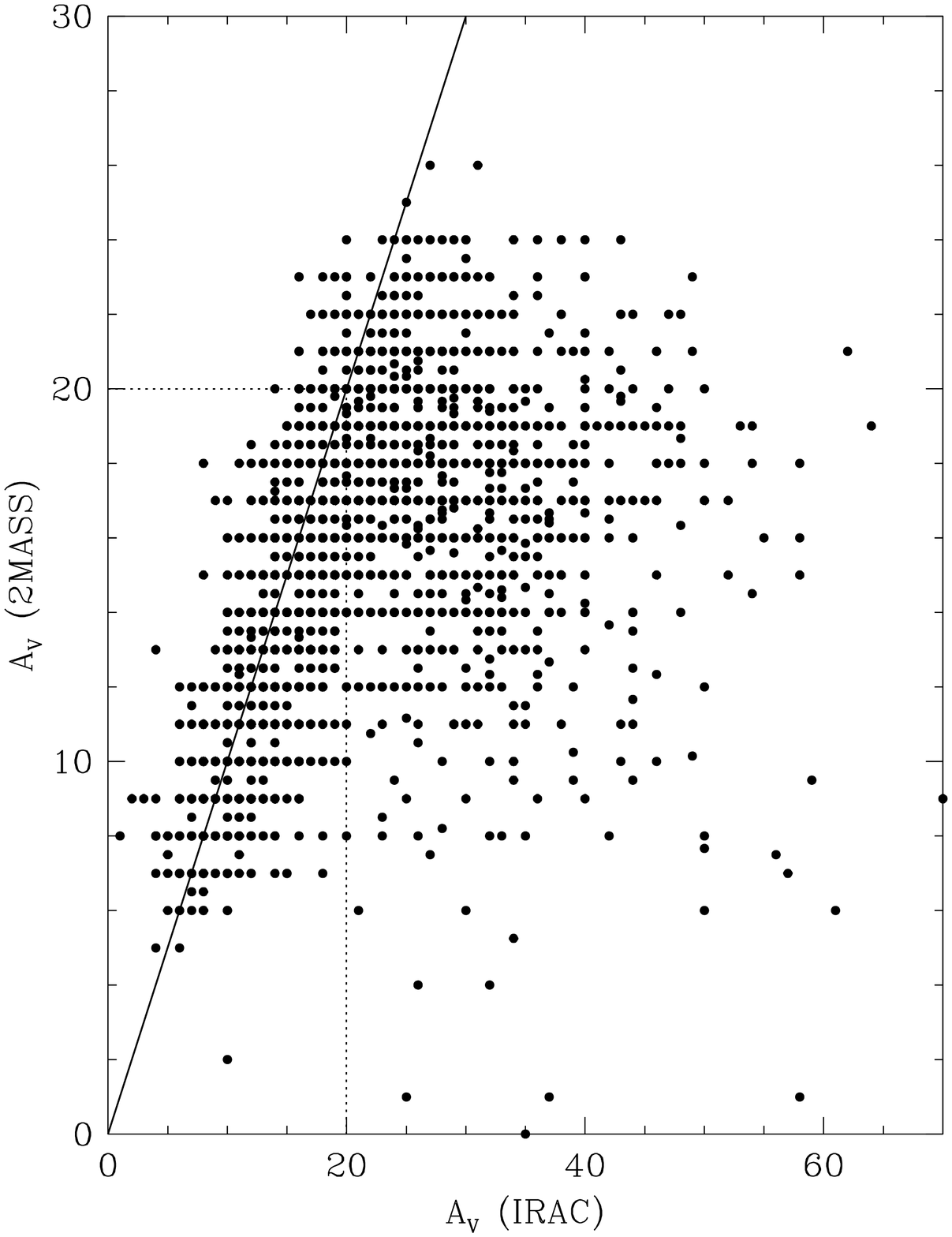}  
%\vspace{3cm} 
%\label{cmd}  
\caption{Extinction values derived from 2MASS compared to those derived from IRAC colours. The straight line gives the identity relation. The dotted line  
indicates the maximum value of $A_{V}$ to which 2MASS is sensitive.}  
\label{diffAv}  
 \end{figure}

Since the $[3.6]$ -- $[5.8]$ colour is less than $J$ -- $K_S$, for a given
amount of reddening the precision it yields in $A_V$ is not as high.  The
loss of accuracy at low $A_V$ can, however, be avoided by using the
near-infrared derived values for mapping when they are available.  At larger
$A_V$, the results from the $[3.6]$ and $[5.8]$ map are, of course, to be
preferred.

 \begin{figure*}[ht!]  
\centering  
\includegraphics[angle=0,width=6cm]{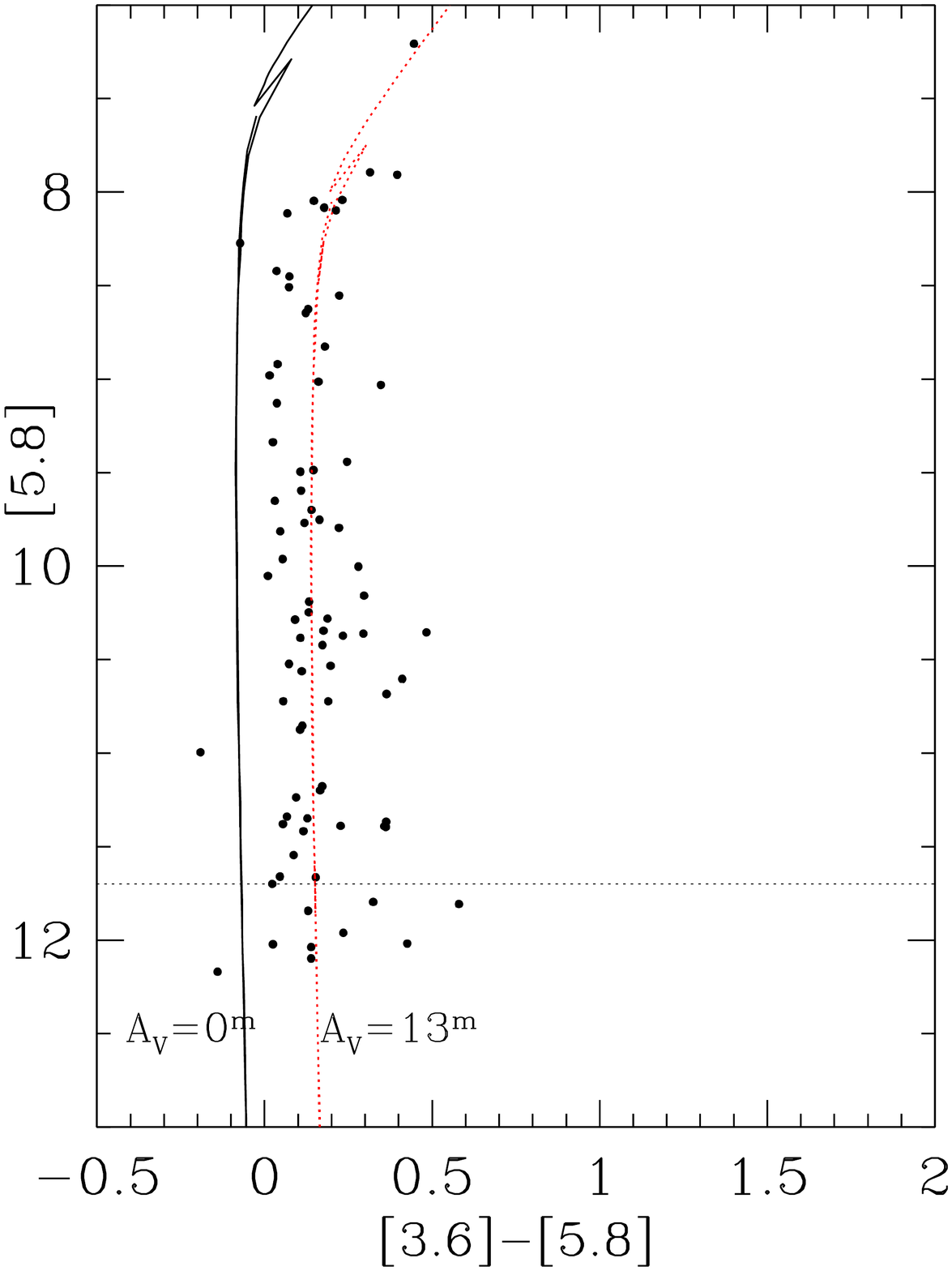}  
\includegraphics[angle=0,width=6cm]{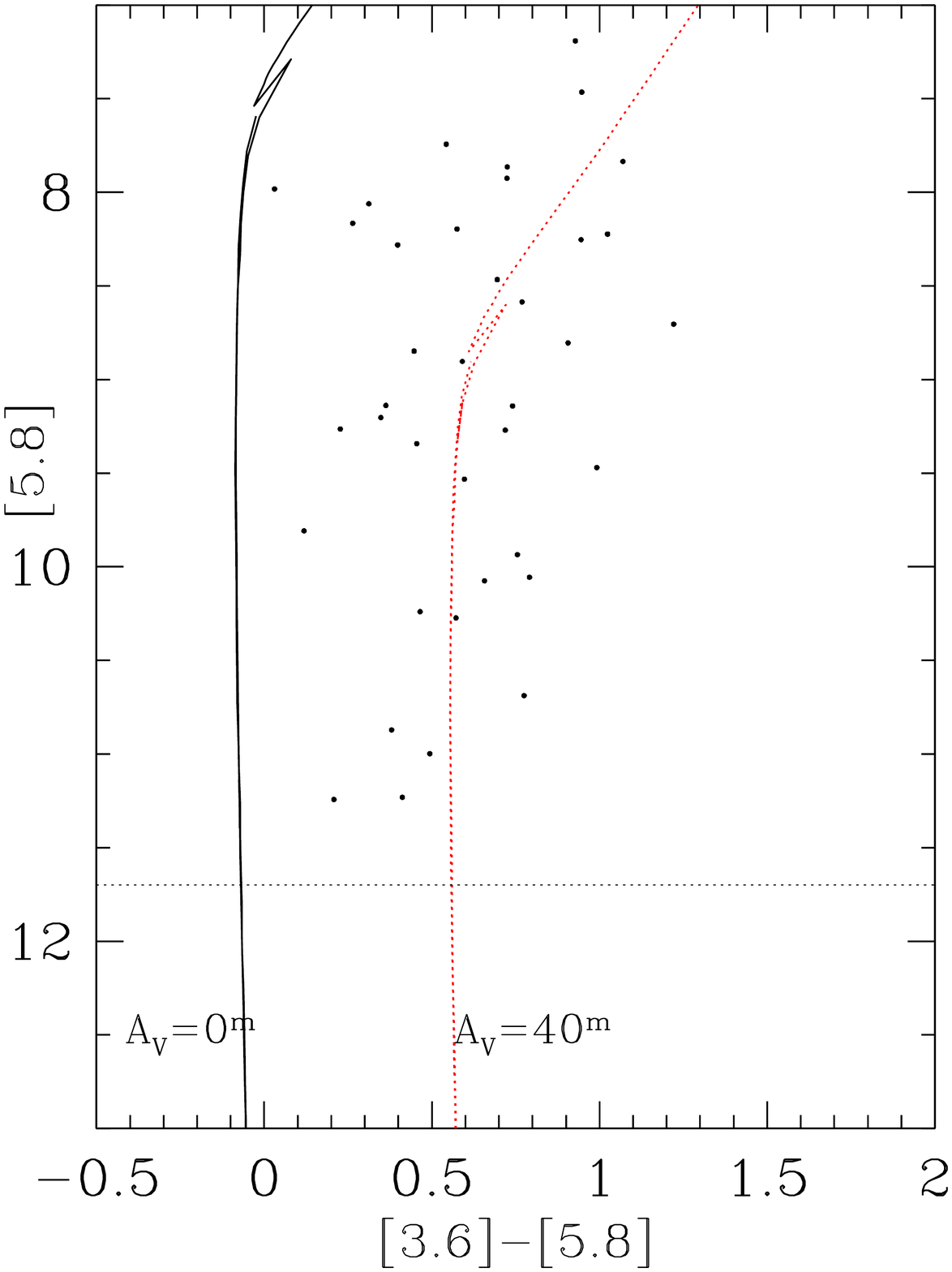}  
\includegraphics[angle=0,width=6cm]{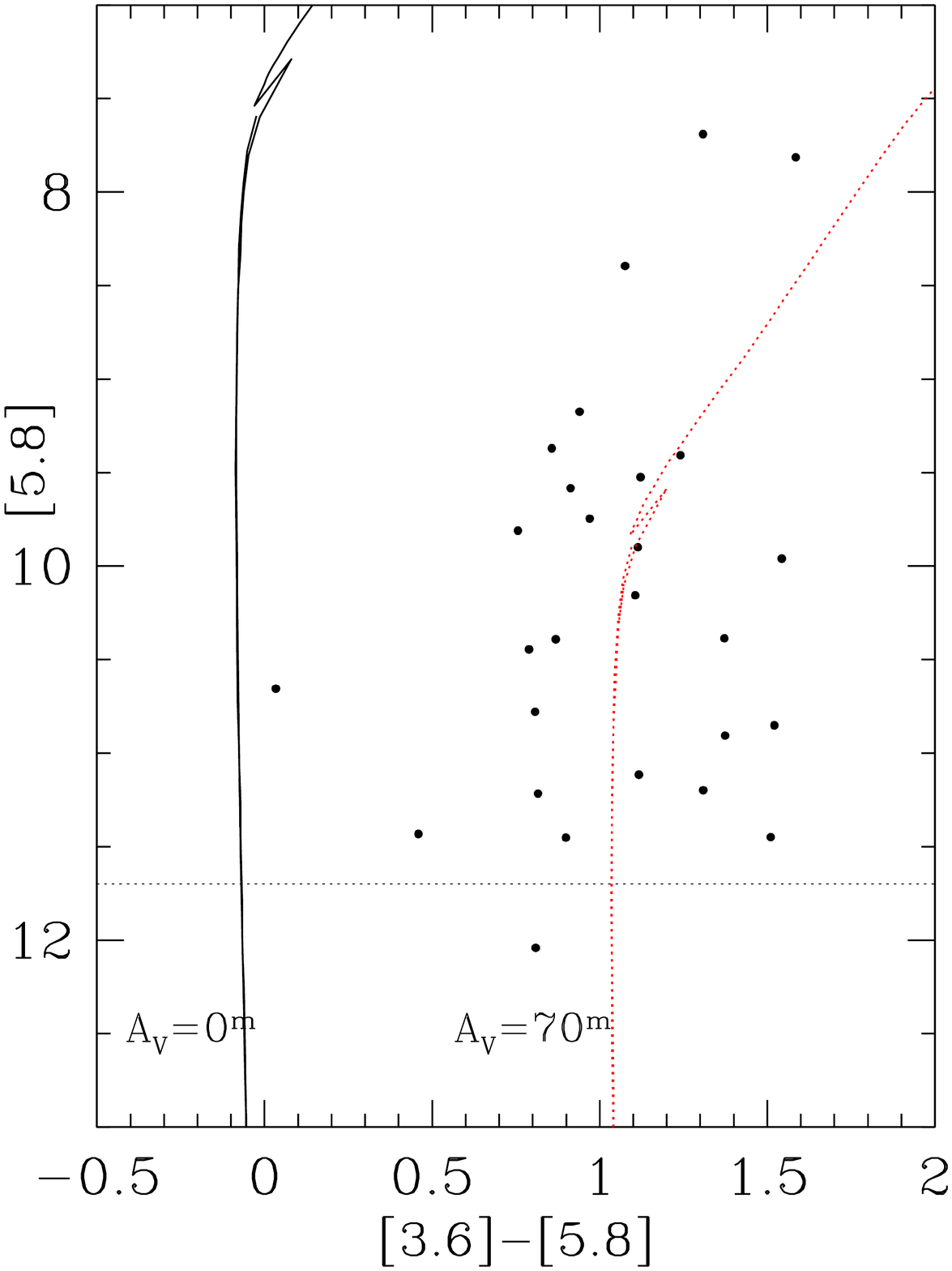} 
%\label{cmd}  
%\vspace{3cm} 
 
\caption{ [5.8] vs. [3.6]$-$[5.8] colour-magnitude diagrams  towards regions of 
low (left panel), intermediate (middle panel) and high (right panel) 
extinction (see text) in the GALCEN survey. The corresponding isochrones are 
displayed. The horizontal dotted line  indicates the 
approximate confusion limit. The corresponding fields (from left to right) 
are $\alpha$ = 265.93596$\deg$, $\delta$ = -28.44771$\deg$; $\alpha$ = 
266.16275$\deg$, $\delta$ = -28.44771$\deg$; and $\alpha$ = 
267.027222$\deg$, $\delta$ = -28.967888$\deg$. The size of each field is 
2\arcmin. The region with $A_{V} = 70^{m}$ is located at $\alpha$ =267.027222$\deg$ and $\delta$ = -28.967888$\deg$ and is associated with the infrared dark cloud IRDN3
of Dutra \& Bica (2001) who list it as an opaque region at $K_{s}$ with an
angular diameter of 3\arcmin.}
 
\label{cmd}  
\end{figure*}

Different ages or metallicities for the Bulge population do not
significantly affect the mid-infrared colours of RGB stars, as will be seen. 
The sequences in the sampling fields are well defined in the case of low
extinction while there is some scatter when it is high.  This scatter arises
from variable extinction on small scales within the Bulge and along the line
of sight.  Towards regions with very high extinction ($A_V$
$>$ 60), the number densities in the CMD can drop significantly due to the loss of the fainter stars. Such points have relatively large errors.

Figure \ref{diffAv} compares the extinction values derived from $J$ and $K_s$
2MASS data and those derived from the $[3.6]-[5.8]$ IRAC colours. There is a
tendency towards a linear relation between $A_{V}$(2MASS) and $A_{V}$(IRAC)
up to $A_{V} = 20$ with a dispersion of $A_{V}$ $\simeq$ 3. For higher
$A_{V}$ values, 2MASS can no longer detect the most reddened stars at the GC and is biased towards less reddened foreground stars. For those regions where 2MASS gives low $A_{V}$ values and IRAC high ones,
different optical depths are obtained. The values of $A_{V}$ derived from
$J-K_s$ are representative of the foreground population rather than
the regions hidden by high extinction, while the IRAC bands relate to stars
that are deeper within the clouds.

Figure \ref{cmd} shows the CMDs towards three regions of different
extinction in the GALCEN survey. While for the low extinction field (left
panel) the red giant branch is well defined and the dispersion is low, the
CMD is more scattered in the high extinction regions. Despite the high
extinction, however, we get a sufficient number of sources to determine
$A_V$.

\subsection{The combined extinction map}

Figure \ref{exterror} shows the combined extinction (2MASS and IRAC) map as
well as an intrinsic uncertainty map derived from the isochrone fitting.
As explained before, for $A_{V}(IRAC) < 20$ the 2MASS $J-K_s$ colour was used, while for $A_{V}(IRAC) > 20$ the IRAC [3.6]$-$[5.8] colour was used. 
 The highest $A_{V}$ values we can detect correspond to about
$A_{V} \simeq 90$. These are mostly associated with clouds that are dark
even in the infrared.
 
We will provide in electronic form (at CDS) the three extinction maps as 
well as the error map in the form of individual FITS files. 
  
\subsection{The uncertainty map}

The intrinsic uncertainty map corresponding to the combined interstellar
extinction map is shown in Fig. \ref{exterror} (lower panel). The
uncertainty map is calculated as the standard deviation of the $A_{V}$
distribution.  For $\rm A_{V} < 20$ the intrinsic uncertainty is about 2-3 \,mag while for $\rm A_{V} > 20$ it can go up to 6-7 \,mag. Figure \ref{diffAvgalcen} shows the comparison between our $\rm A_{V}$ determinations based on GALCEN and on the
GLIMPSE-II catalogue (see also Sect. ~2.2). Despite the differences found in
the photometry between these two pipelines, no systematic offset is found
between $\rm A_{V}$ derived from GALCEN and $\rm A_{V}$ derived from
GLIMPSE-II. The typical r.m.s standard deviation between GALCEN and
GLIMPSE-II is 6\,mag in $\rm A_{V}$ which is the overall uncertainty in the
derived extinction map.  For very high $A_{V}$ ($A_{V}$ $>$ 40), the differences between GALCEN and GLIMPSE
become larger. Howver, for those regions the CMD is less well defined and the intrinsic uncertainty increases (see 
Fig. \ref{exterror}).

It should be noted that Nishiyama et al. (\cite{Nishiyama06}) found that the
behaviour of near-IR extinction varies slightly though significantly from
one direction to another near the GC so that there is no universally valid
extinction law for the region. These results were confirmed by Bandyopadhyay
et al.  (\cite{Bando08}). Variations with direction of $A_{\lambda}/A_{V}$ in the
IRAC bands do not appear to be significant at the accuracy that is currently
available, at least in the fields studied by Indebetouw et al (2005) and
Ganesh et al. (in prepration).

\begin{figure*}[ht] 
\centering  
\includegraphics[angle=0,height=9.5cm]{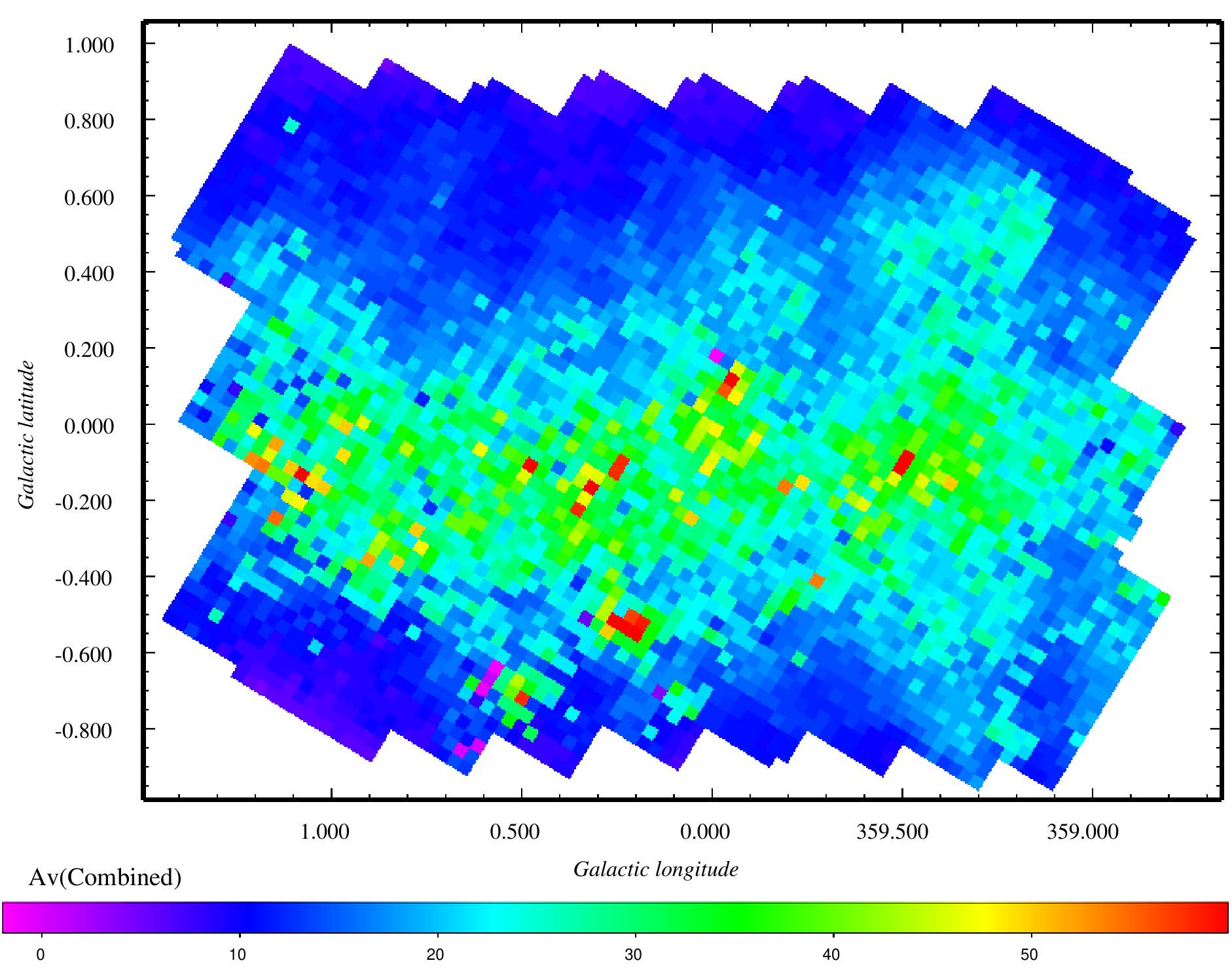}  
\includegraphics[angle=0,height=9.5cm]{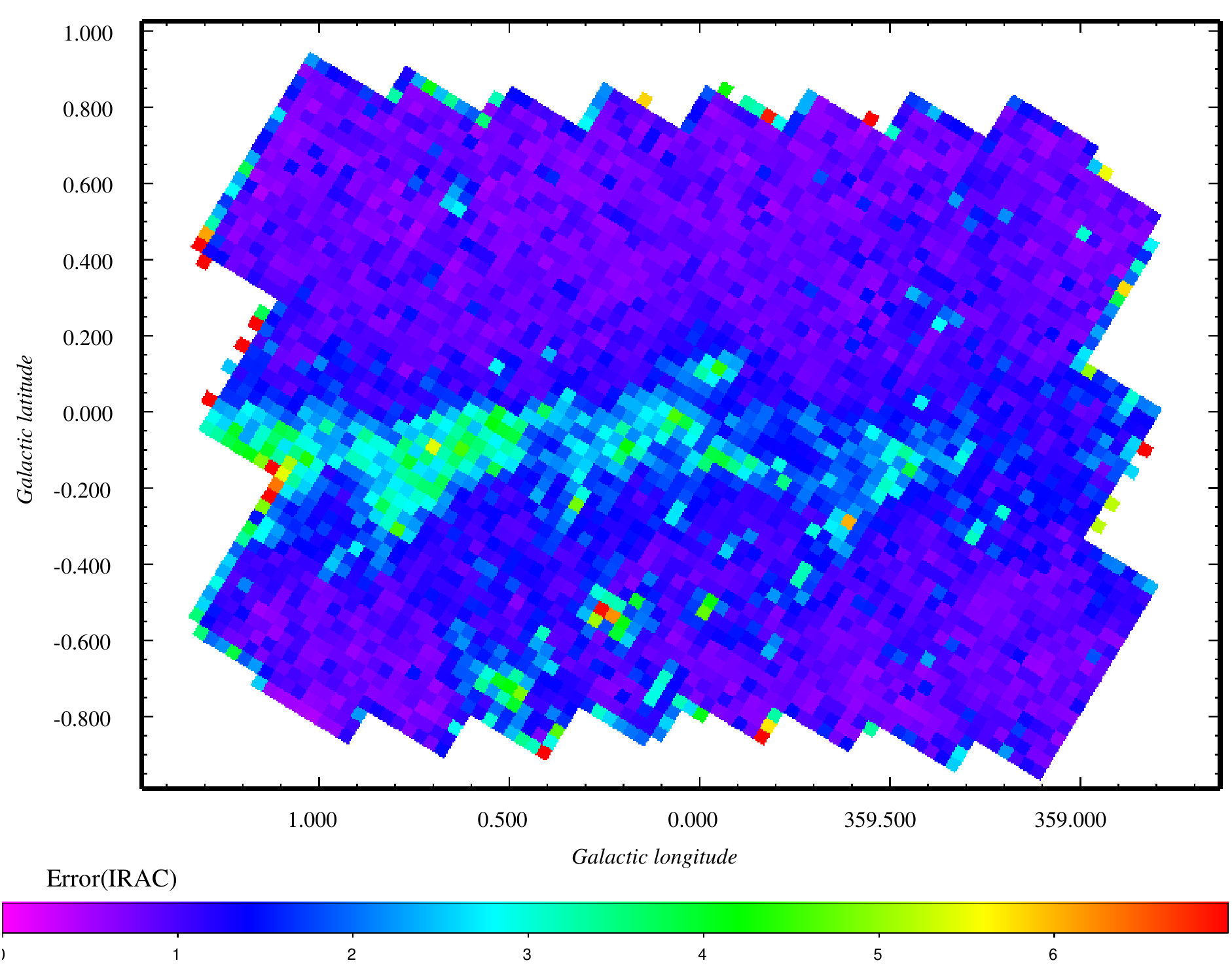}  
\caption{{\em {Upper panel:}} Combined interstellar extinction map of the 
central 280 $\times$ 200\,pc of the Galactic Center.  For $A_{V}(IRAC) < 20$ 
the 2MASS $J-K_s$ colour was used, while for $A_{V}(IRAC) > 20$ the IRAC 
[3.6]$-$[5.8] colour was used.  {\em{Lower panel:}} Intrinsic error map of 
the combined interstellar extinction map.  The error is the sigma of the 
$A_{V}$ distribution.} 
 
\label{exterror}  
\end{figure*}

 \begin{figure}[h] 
\centering  
\includegraphics[angle=0,width=7cm]{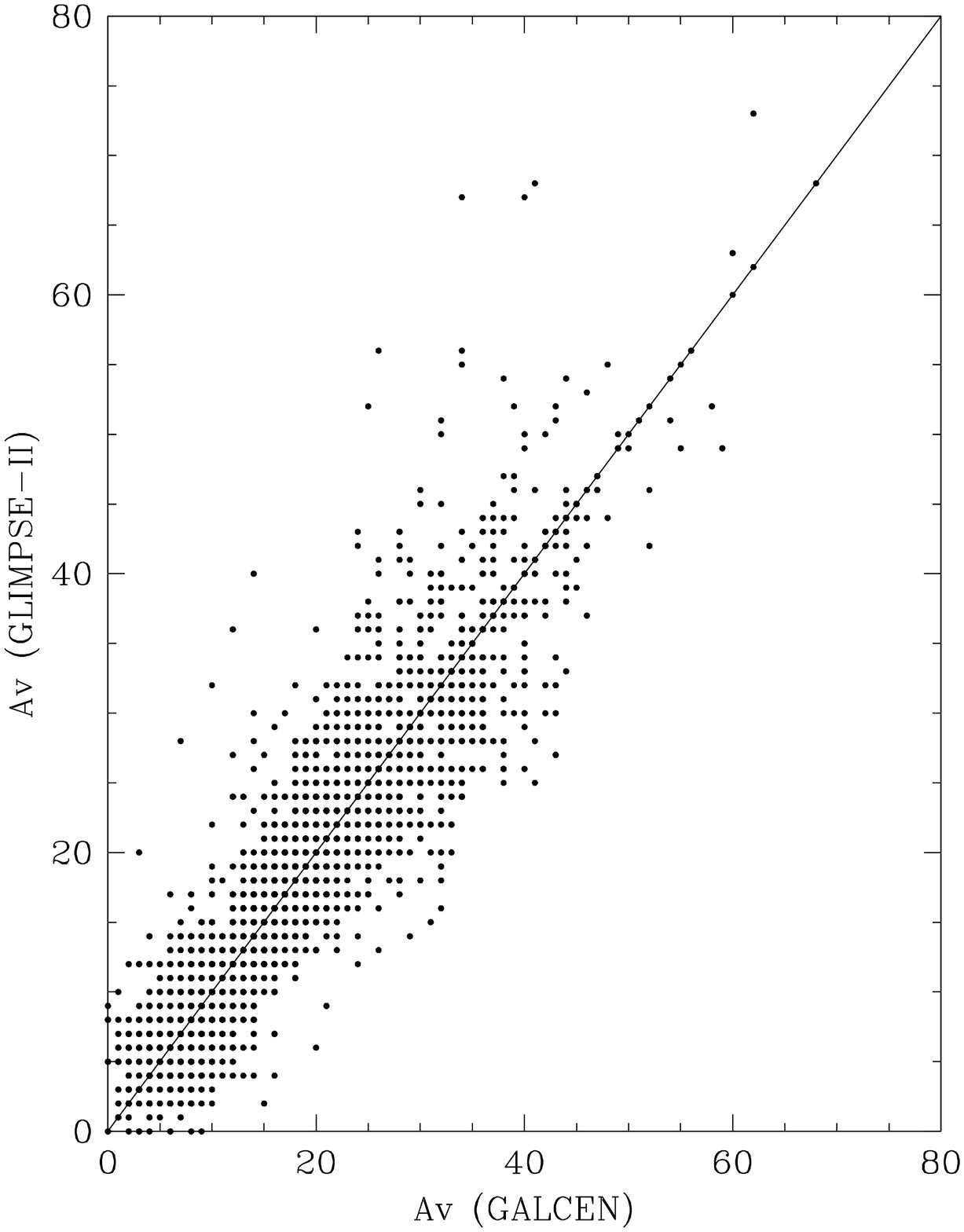}  
%\vspace{3cm} 
%\label{cmd}  
\caption{Extinction values derived from GALCEN  compared to those derived from GLIMPSE-II. The straight line gives the identity relation.}  
\label{diffAvgalcen}  
\end{figure}

Gosling et al. (\cite{Gosling06}) found complex small-scale structure in the
infrared extinction towards the GC, with a typical size of 5--15\arcsec,
corresponding to 0.2--0.6 pc. They demonstrated that significant granularity
appears to be present in high and variable extinction fields, indicating
that extinction at $J$ and $H$ may have significantly higher values on
smaller scales than those previously derived for the GC as a whole. 
%However,
%they state that the effect is low or absent at $K$ and we have presumed that
%this is also the case at longer wavelengths.

\section{The Long Period Variables near the Galactic Center}  
 
In the remainder of this work we discuss the long-period variables of the
Galactic Center region. Figure \ref{ImageLPV} shows the 2MASS image of the
GC region at $K_{S}$. The known LPVs of Glass et al. (2001) are superimposed
and are bright in $K_{S}$. We note that patchy and filamentary dark clouds (appearing white in Fig. 7) often vary on size scales smaller than the 2\arcmin sampling boxes used here to calculate the extinction. This can contribute to the photometric errors in the extinction correction towards individual sources.
%Note the strong, patchy and occasionally
%filamentary extinction which, in spite of the small size of the sampling
%boxes used to calculate the reddenings, may still contribute to the
%photometric errors after correction.

\subsection{Stellar isochrones}  
  
The newest set of isochrones for the RGB/AGB phase include  an improved
treatment of the thermally pulsing asymptotic giant branch (TP-AGB) phase
(Marigo et al. \cite{Marigo2007}).  These isochrones predict realistic
tracks for the TP-AGB, taking into account mass-loss and pulsation. The
whole of TP-AGB evolution is now treated in a realistic way, especially the
crucial effects of the third dredge-up, hot bottom burning and variable
molecular opacities.  

The dust models incorporated in these isochrones are an extension of the
work by Groenewegen (\cite{Groenewegen06}). They have been calculated with a
1-dimensional dust radiative transfer code that solves the radiative
transfer equation and the thermal balance equation in a self-consistent way.

The main physical inputs to the model are the luminosity, distance,
photospheric spectrum, mass-loss rate, dust-to-gas ratio, expansion
velocity, dust condensation temperature and composition of the dust. For
oxygen-rich stars two main species of dust are considered: aluminium oxide
($\rm Al_{2}O_{3}$) and silicate dust. Blommaert et al. 
(\cite{Blommaert06}) studied the CVF ISOCAM spectra of a sample of AGB stars
with low mass-loss in the Galactic bulge and found that the dust content is
dominated by $\rm Al_{2}O_{3}$ grains. 

Isochrones for different metallicities, Z=0.008, Z=0.019 and Z=0.038, and
ages between 0.5 and 10 Gyr, have been calculated for the Spitzer IRAC
filters. In the present work, we consider the isochrones placed at the
distance to the GC (8.0 kpc, Reid \cite{Reid1993}).

\begin{figure}[h]  
\centering  
\includegraphics[angle=0,width=7.5cm]{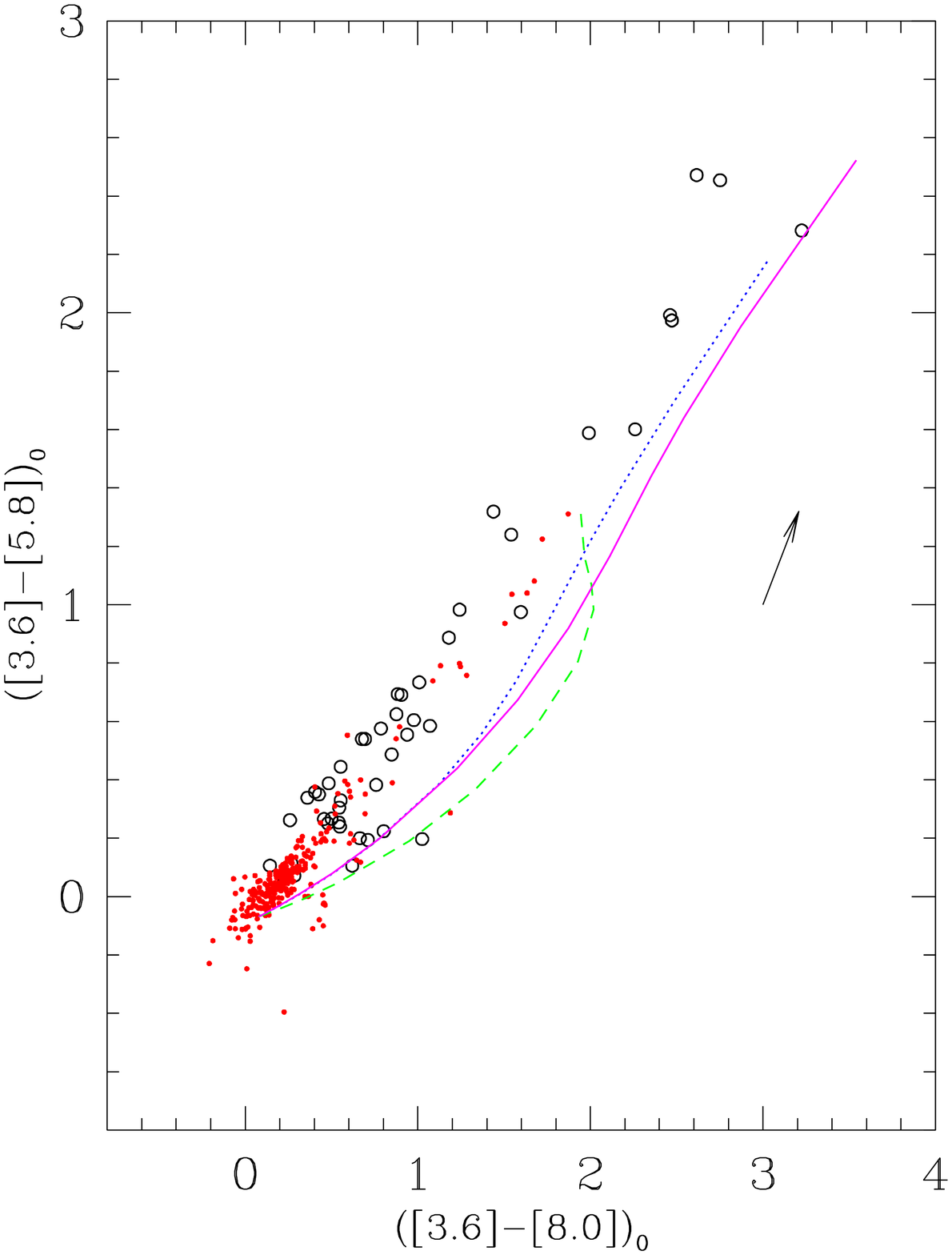}   
%\label{ccd1}  
%\vspace{3cm} 
 
\caption{$\rm ([3.6]-[8.0])_{0}$ vs $\rm ([3.6]-[5.8])_{0}$ diagram. Open 
circles indicate the % non-saturated  
GC LPVs of Glass et al. (2001).  
Filled red circles denote oxygen-rich LPVs in the LMC (see text).  The straight 
lines are models computed by Groenewegen (2006) using realistic stellar 
atmosphere models.  The dashed green line  is a model with 100\% aluminium oxide dust, the dotted blue line 
is 60\% aluminium oxide and 40\% silicate, and the solid magenta line is 100\% silicate 
dust. The extinction vector of $A_{V}=20$ is indicated.} 
 
\label{ccd1}  
\end{figure}

\subsection{LPVs in the Spitzer colour-magnitude and colour-colour diagrams}

\begin{figure*}[htbp]  
\centering  
\includegraphics[angle=0,width=11.0cm]{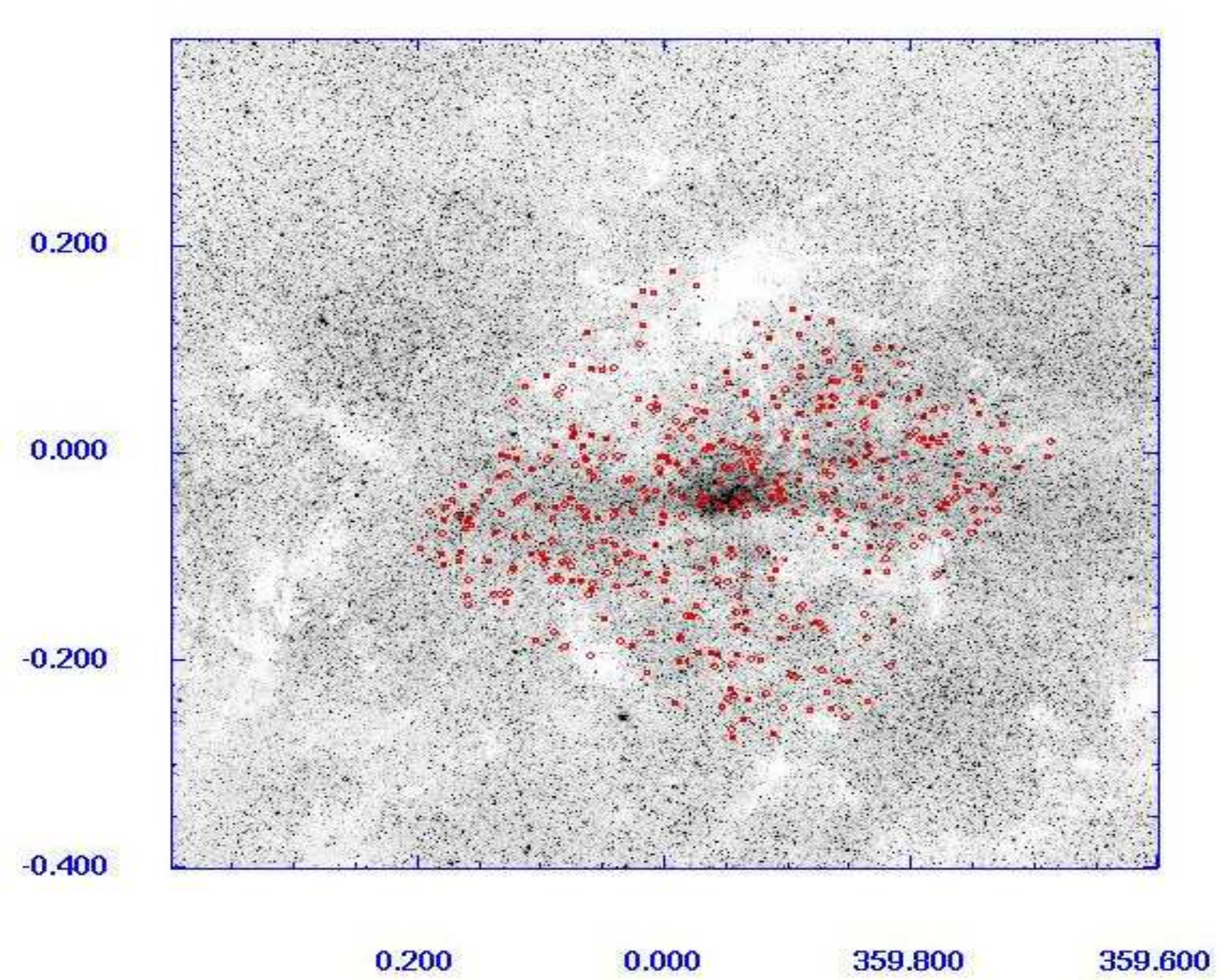}  
\caption{2MASS image (reversed grayscale) of the Galactic Center region at $K_{S}$. The known  
LPVs from Glass et al. (2001) are superimposed with red circles.}  
\label{ImageLPV}  
\end{figure*}  
 
% \vspace*{-1cm}
 
\begin{figure*}  
\centering  
\includegraphics[angle=0,width=6.8cm]{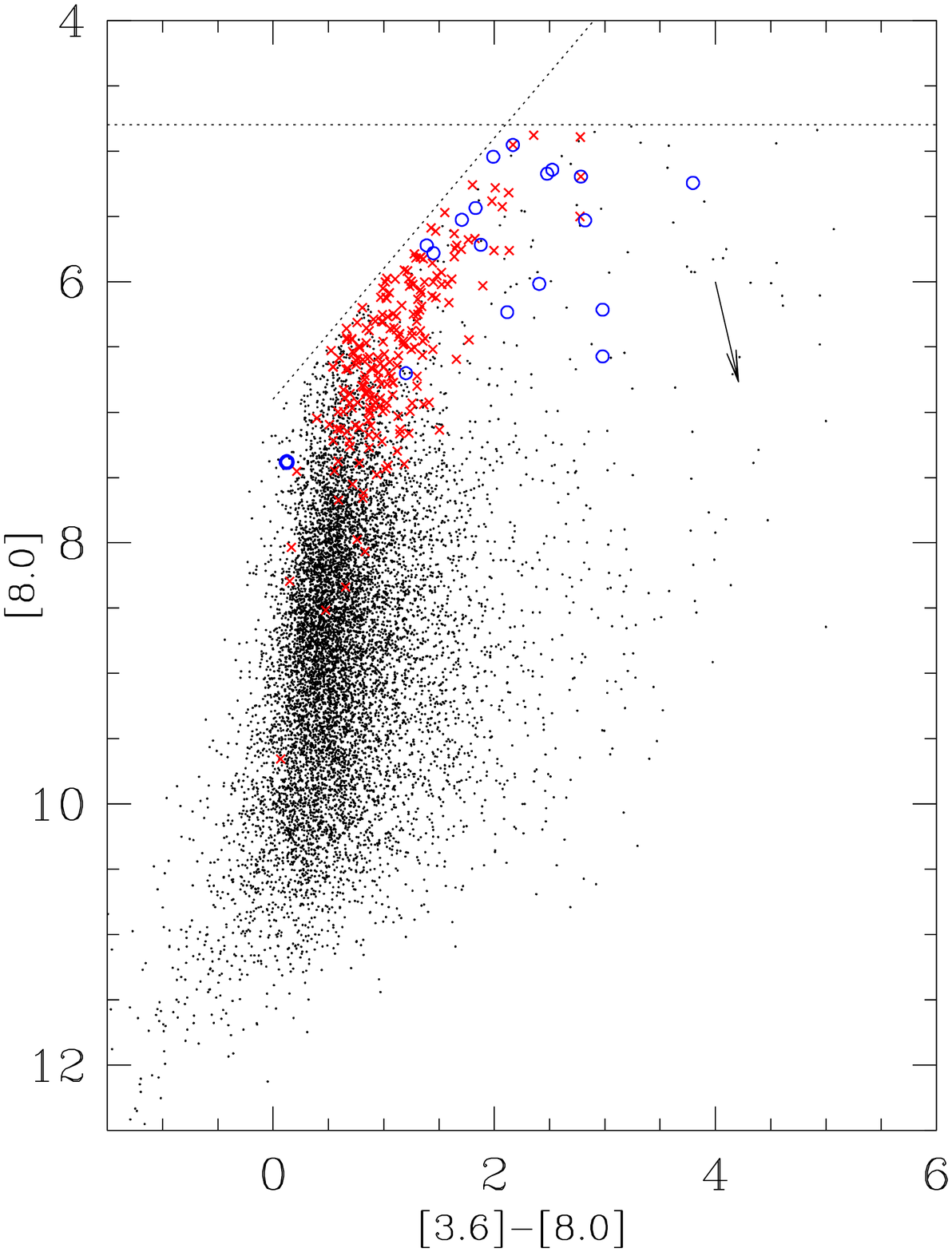}  
\includegraphics[angle=0,width=6.8cm]{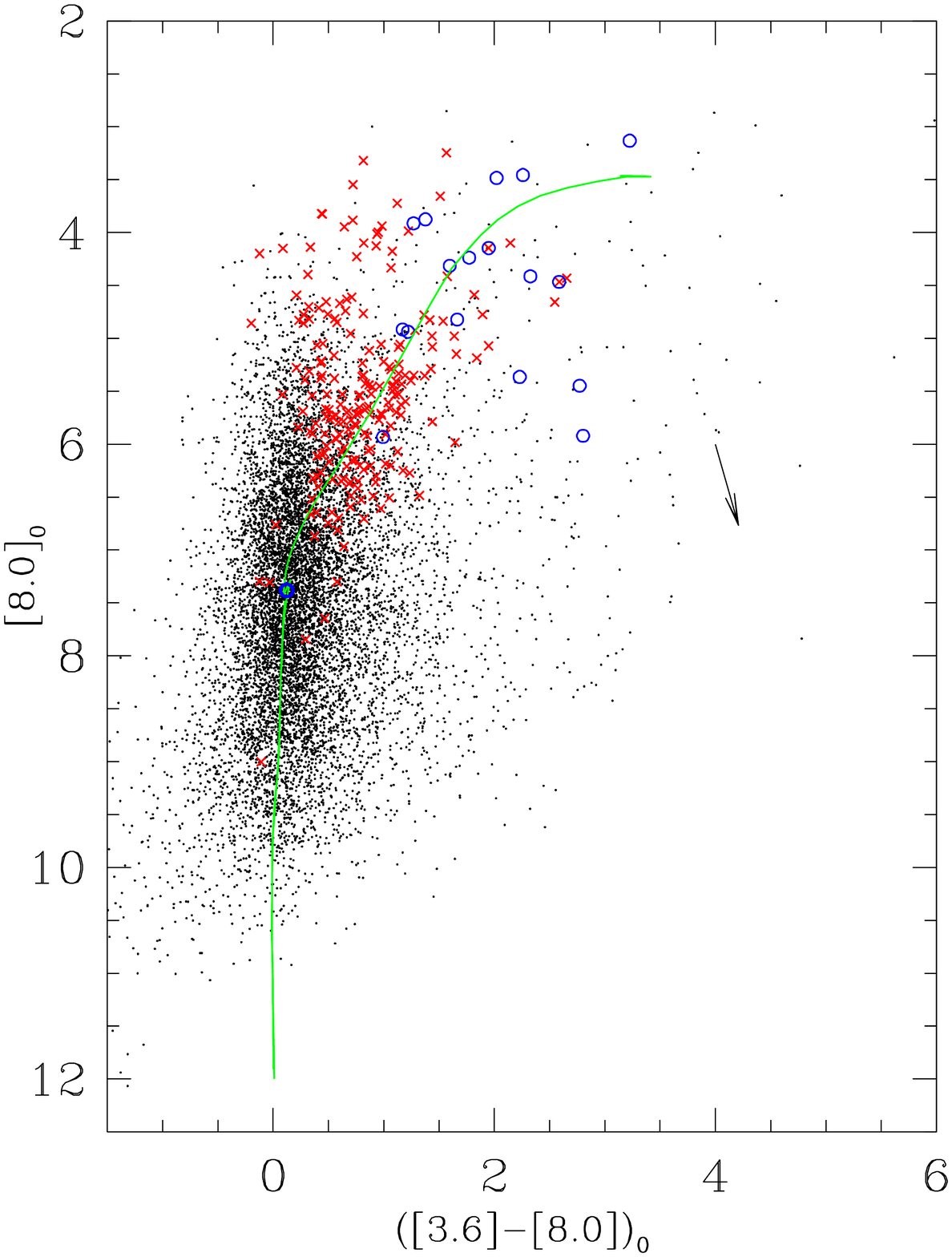}  
%\vspace{3cm} 
 
\caption{{\em Left panel:} [3.6]$-$[8.0] vs [8.0] colour-magnitude diagram  of 
the GALCEN field,  with the known LPVs (red crosses) and OH/IR stars 
(blue open circles) superimposed. The extinction vector  for $A_V$=20 is 
indicated. The dotted 
horizontal line indicates the saturation limit at [8.0] while the dotted 
diagonal line shows the saturation limit at [3.6].  {\em Right panel}: Same 
diagram but dereddened (see text). Superimposed is the isochrone for 10 Gyr 
and Z=0.02 put at a distance of 8\,kpc.} 
 
\label{ch14_ch4}  
 \end{figure*}  
   
Figure \ref{ch14_ch4} (left panel)  shows  the [3.6]$-$[8.0]   vs [8.0]
diagram. Indicated are also the approximate saturation limits at [3.6] and
[8.0].  In total, only 248 known LPVs have reliable (unsaturated) IRAC
photometry. On the right panel of Fig.~\ref{ch14_ch4}, the dereddened
[3.6]$-$[8.0] vs [8.0] diagram, derived using the extinctions given in
Fig.~\ref{exterror}, shows clearly that the locus of the known AGB stars is
above the RGB tip ($\rm [8]_{0} \simeq 7.3$). As can be seen, the isochrones
of Marigo et al. (2008) are a good fit to the dereddened CMD.  This
agreement confirms the correctness of the derived extinctions.  The
locations of the isochrones change only marginally with age and/or
metallicity (e.g. from 1 to 10 Gyr and $\rm 0.008 < Z < 0.03$ in these
bands. The AGB stars show a large scatter in this diagram due to strong
variability (we only have single-epoch measurements) and also to the
infrared excesses arising from heavy mass-loss (see Sect. \ref{massloss}). 
A similar diagram can be drawn for LPVs at [4.5] where we find only 84 AGB
variables to be unsaturated.

\subsubsection{Comparison with Spitzer observations of LMC LPVs}
  
Cioni et al. (\cite{Cioni01}) examined a field of about 0.5 square degrees
around the optical center of the Large Magellanic Cloud.  Light curves in the EROS V and R bands were
obtained for 334 variables, most of which are of long period (including
Miras and semiregulars). In addition they obtained spectroscopy of about 100
AGB variables whose visible spectra were classified as carbon- or
oxygen-rich. We used their sample of known AGB stars and cross-identified
their catalogue with the SAGE (Surveying the Agents of a Galaxy's Evolution)
MIPS and IRAC imaging survey (Meixner et al. \cite{Meixner06}). In the
latter work, a field of 7 sq degrees of the LMC was surveyed between 3.6 and
160 \,$\mu$m. In total, 331 objects were found in the
SAGE catalogue.  For further discussion, we will refer to this as the
LMC-AGB sample.
  
Figure \ref{ccd1} shows the dereddened [3.6]$-$[8.0] vs [3.6]$-$[5.8]
colour-colour diagram for both the Galactic Center and the LMC AGB LPVs.  We
have superimposed for comparison theoretical dust models by Groenewegen
(\cite{Groenewegen06}). These utilise several types of dust for the O-rich
AGB stars: 100\% aluminium oxide, a combination of 60\% aluminium oxide and
40\% silicate, and pure silicate dust (see Groenewegen 2006 for a
discussion).  All three models fail to match the observed colours and a significant
offset is clearly visible.
%With the exception of the model with pure aluminium oxide
%(green line), the dust models reproduce qualitatively the observed colours:\
%but with a significant offset in [3.6]$-$[5.8].
  
Our comparison is restricted to oxygen-rich AGB stars (283 stars). It should
be noted that we split the LMC-AGB sample into oxygen-rich and carbon-rich
AGB stars using only the $J-K$ colour.  Though Cioni et al. (\cite{Cioni01})
showed that carbon stars can in general be eliminated by avoiding colours in
the range $1.4 < J-K_s < 2.0$, they warn (see their Fig. 3) that stars
selected in this way will contain a certain fraction of carbon stars.
 
Figure \ref{ccd1} shows that the GALCEN variables and LMC oxygen-rich
variables only partly overlap in the colour-colour diagram. The GALCEN LPVs
show much redder colours in [3.6]$-$[8.0] and [3.6]$-$[5.8] than the LMC
sample. These stars also have a higher proportion of long periods
(Groenewegen \& Blommaert 2005) relative to the LMC and the remainder of the
Galactic Bulge. The redder colours in the IRAC bands may indicate that
mass-loss rates of AGB stars in the GC are higher in general than in the
LMC. However, the differing circumstellar dust-to-gas ratios between the LMC and GC makes
direct comparison difficult. The differences can also be a manifestation of
the strong relation between period and mass-loss (see Sect. \ref{massloss}).

\section{The Period-magnitude relations}  
  
AGB variable stars in the Galactic Bulge follow  a well-defined
period-magnitude relation in the near infrared. Glass et al. (2001) showed
that the mean $K$ magnitude for the Glass-LPVs, measured over a complete
cycle, showed some tendency to follow a log $P$ vs. $K$ relation in spite of a very high dispersion due to the large and unknown variations in $\rm A_{V}$. Some of the scatter may also have arisen because of crowding effects or insufficient observations having
led to incorrect periods, especially at the long (log $P > 2.8$)
and short ends (log $P < 2.3$) of the range. The Glass et al. (2001) sample  contains a higher fraction of  long periods than e.g. the SgrI field. Thus, circumstellar reddening due to heavy mass-loss  affects the  log $P$ vs. $K$ relation by depressing the emergent fluxes at longer wavelengths.

Figure ~\ref{PL} shows the IRAC magnitude vs. log $P$ relations for the
Glass-LPVs, after correction of the photometry for interstellar extinction.
We find following relations for the Glass-LPV sample: \\
 \noindent
\noindent
 $[3.6]$  =  log P $\times$  -5.21 +18.94\\ 
\noindent
  $[4.5]$  =  log P $\times$  -5.71 +19.98\\
  $[5.8]$  =  log P $\times$  -6.10 +20.70\\
  $[8.0]$  =  log P $\times$  -6.73 +21.96\\

The typical r.m.s deviation for the four [IRAC]  bands are  0.71 , 0.62, 0.60 and 0.61 mag in the period range between $\rm 2.2 < log\,P < 2.8$.  This is smaller than  at  $K_{0}$  where the r.m.s is 0.95\,mag (corrected according to the extinctions determined in this work) as the IRAC bands
are less affected by any residual errors in the extinction. Even within the
IRAC bands, the dispersions at the longer wavelengths are less than that at 3.6 $\mu$m.  The Glass-LPVs with periods greater than 900 days do not follow the log P vs. [IRAC] relations.
 Their periods  could be incorrect as these only have a few observations  (see Glass et al. 2001) and in addition  these stars suffer heavy mass-loss which depress their [IRAC] fluxes.

% The
%dispersion is smaller by a factor of 2  than at $K_{0}$ (corrected according to the extinctions determined in this work) as the IRAC bands
%are less affected by any residual errors in the extinction.  Even within the
%IRAC bands, the dispersion decreases at longer wavelengths as the effects of
%extinction decrease. 

Superimposed on Fig.~\ref{PL} is the LMC-AGB star sample. Again, we have
chosen only oxygen-rich Mira variables in order to make direct comparisons
with the Glass-LPV sample. We have put the LMC LPVs at the distance of the
GC using a distance modulus difference of $\Delta m = 4.0$ mag. Within the errors, the Glass-LPVs do follow the LMC relation.

\begin{figure*}  
\centering  
\includegraphics[angle=0,width=14cm]{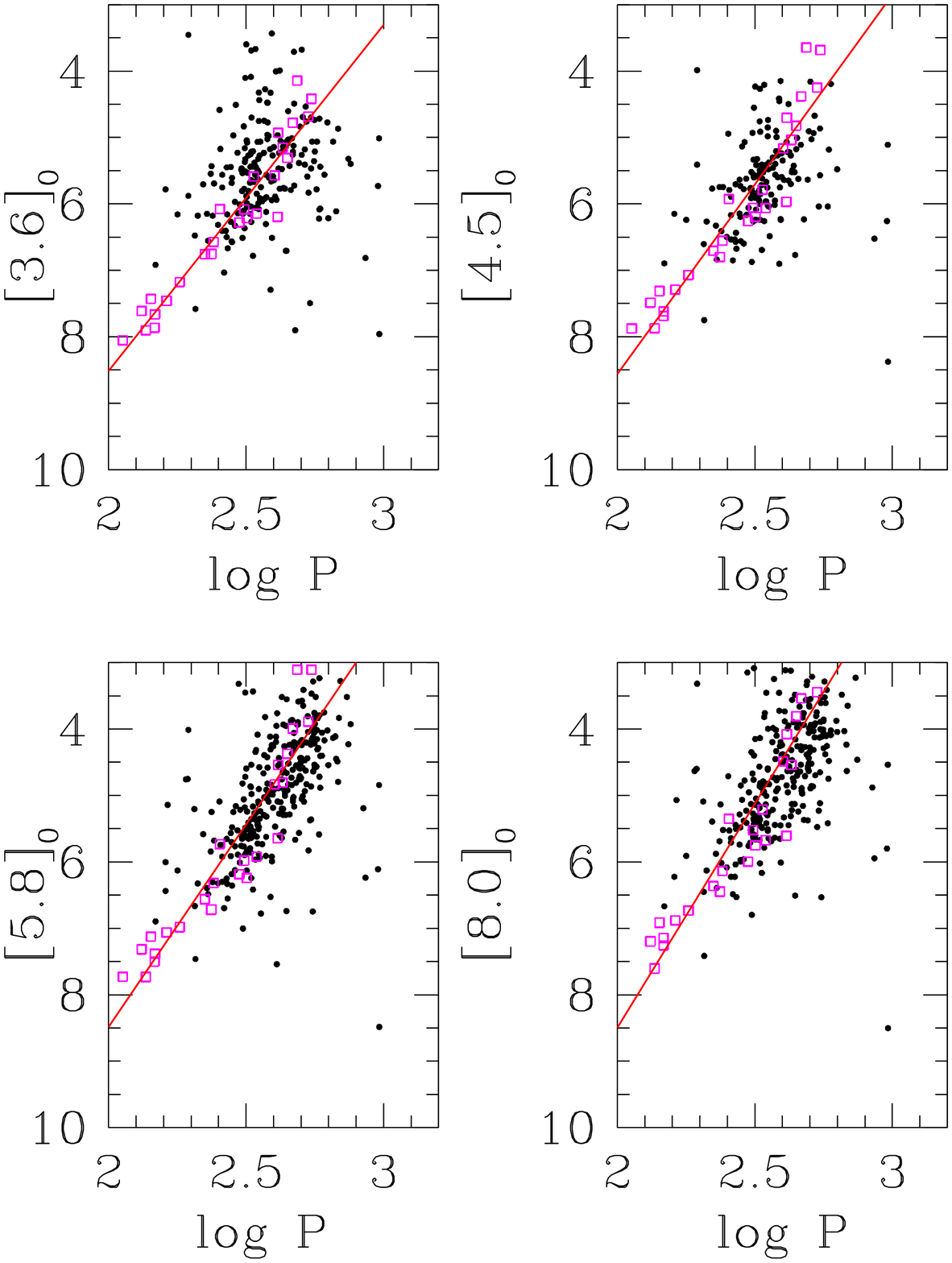}  
%\vspace{3cm}  
 
\caption{Dereddened IRAC magnitudes (using the extinction map as described
in Sect. 3) vs. log $P$ for LPVs in the Glass et al. (2001) GC field
(Glass-LPVs; black dots), where $P$ is the period.  Open magenta squares are
LMC-AGB stars (oxygen-rich Mira variables). The straight red line is a
least-squares fit to the LMC-AGB data set.}
 
\label{PL}  
\end{figure*}  
  
As mentioned, the scatter in the period-magnitude relations for 
the Glass-LPVs is much higher than in the LMC-AGB sample. This is due to the 
uncertainties in the extinction determination. No attempt 
has been made to compensate for the triaxial distribution of stars in the 
Bulge. Recent studies of the central region of our Galaxy suggest a smaller 
bar-like structure within the main bar (see e.g. Alard \cite{Alard2001}, 
Nishiyama et al. \cite{Nishiyama05}) which might be partly responsible for 
the observed scatter. However, the parameters of the central bar (such 
as its angle, length and  orientation) are still poorly known. 
  
In contrast to the light curves obtained by Glass et al. (2001), we have 
only single epoch measurements in the IRAC bands. The $K$ amplitude 
distribution of the Glass-LPVs peaks around 0.8\,mag, going up to 1.5\,mag 
for the OH/IR stars. In order to estimate the expected amplitude of LPVs in 
the IRAC filter bands, we used the calculated colours of Galactic 
oxygen-rich AGB stars by Marengo et al. (\cite{Marengo06}). Several such 
objects were observed by ISO at different epochs.  Mira variables, such as Z 
Cyg, can show variations up to 1\,mag in each IRAC channel, while in the 
most extreme cases, such as the dust-enshrouded OH/IR star V1300 Aql, 
variations of 1.4\,mag may occur in each filter. 
  
Up to now, period-magnitude relations have been mainly studied in the 
intermediate Bulge (such as Baade's Window) and the LMC, where the log $P$ vs 
$K_{S}$ relations seem to be independent of the galactic 
environment (see e.g. Schultheis et al. \cite{Schultheis2004}, Glass \& van
Leeuwen \cite{Glass07}). Groenewegen \& Blommaert (\cite{Groenewegen05}), by
using OGLE data, also compared the period-magnitude relations in the Galactic
bulge and the Magellanic Clouds and concluded that there is no difference in
their slopes.  Rejkuba (\cite{Rejkuba04}) found that even Miras in Cen A
follow a similar relation.
 
The Galactic Center, however, has up to now been excluded from this
comparison. In asking whether substantially different behaviour should be
expected we first summarize recent abundance studies in order to place the
GC in context with the Bulge ([Fe/H] $\simeq$ $-0.1$; alpha-enhanced) and
the LMC ([Fe/H] $\simeq$ $-0.4$; no alpha-enhancement).  Studies have shown
that in the GC the mean [Fe/H] is nearly solar; [Fe/H] $\simeq$ 0.1
(Ram\'{i}rez et al. \cite{Ramirez2000}; Najarro et al. \cite{Najarro04},
\cite{Najarro08}; Cunha et al. \cite{Cunha07}). On the other hand, Ram\'{i}rez et al.
(\cite{Ramirez2000}) found that the distribution in metallicity is
significantly narrower than in the Bulge, emphasizing the difference between
its stellar population and that of the GC.  
%Cunha et al.
%(\cite{Cunha07}) and Najarro et al. (\cite{Najarro08}) have measured the
%ratio of alpha elements to iron in M supergiant/AGB stars and Luminous Blue
%Variable stars, respectively.  They both conclude that [$\alpha$/Fe] in GC
%stars is a factor of $\sim$2 higher than in the galactic disk, indicating
%that SNe II have had a larger influence on the GC population.
 
Figure~\ref{PL} shows clearly that the slopes of the period-magnitude
relations at the IRAC wavelengths for the Glass-LPVs are similar to those of the LMC. This suggests
that any dependence of the log $P$ vs. $IRAC$ relations on abundance, if
present at all, must be small. Within the uncertainties, there is no
evidence for a significant difference between the period-magnitude relations
in the LMC and in the GC. This is in agreement with Whitelock et al.
(\cite{Whitelock08}) who found, after reanalyzing published lightcurves of
AGB variables in the LMC, similar zero points in the period-$M_{K}$
relations for systems with different metallicities. They did not, however,
include in their analysis the variables towards the GC, which is thought
to be  the most metal-rich environment. Thus, our analysis shows that the
 apparent universality of the PL relations extends to the IRAC bands.
  
\section{Mass-loss} \label{massloss}  
  
AGB stars contribute more than 70\% towards the enrichment of the dust 
component of the interstellar medium (ISM) in the solar neighbourhood 
(Sedlmayr \cite{Sedlmayr94}) and hence it is important to study their 
mass-loss in other parts of the Galaxy.  One of the most promising tools 
for determining mass-loss rates is the combination of near-IR and mid-IR 
colours such as the IRAS $K_{0}-[12]$ or the ISOGAL $K_{0}-[15]$ 
colours (see e.g. Whitelock et al. 1994; Habing 1996; Le Bertre \& Winters 
1998; Omont et al. 1999; Jeong et al. 2002, Ojha et al. 
\cite{Ojha03}, \cite{Ojha07}). 
 
We used the dust radiative transfer code for oxygen-rich AGB stars from
Groenewegen (2006) to derive theoretical mass-loss rates. The procedures
described by Ojha et al. (\cite{Ojha07}) to derive observed mass-loss rates
based on the $(K_{S} -[15])_{0}$ colour (see Equ. 1).
 
\begin{equation} 
 % log\, dM/dt = -8.617 \times x - 0.0641 \times x^2 + 0.0018083 \times x^3 
\log\dot{M}~=~-8.6171~+~0.85562~x~-0.064143~x^2~+~0.0018083x~^3 
\end{equation} 

\hspace*{-0.7cm} where $ x = (K-[15])_{0}$

As in Ojha et al. (\cite{Ojha07}) we are considering an AGB star model with
$T_{\rm eff}= 2500\, \rm K$ and 100\% silicate composition. Note that the
empirical values determined by Groenewegen (2006) are for mass-loss rates
smaller than 2 $\times$ $10^{-5}$ M$_{\odot}$ yr$^{-1}$.  For the high
mass-loss end we extrapolate the empirical relation.
 
The distribution of  mass-loss rates is displayed in 
Fig.~\ref{mass-loss}. The derived values of $\dot{M}$ range from $\rm 10^{-8}$ to $\rm 10^{-5}$
M$_{\odot}$ yr$^{-1}$. The distribution, however, is definitely incomplete
for $\dot{M}$ $<$ 1 $\times$ $10^{-7}$ M$_{\odot}$ yr$^{-1}$. We 
emphasize that the determination of these rates is very uncertain as
discussed by Schultheis et al. (\cite{Schultheis2003}). The uncertainties
could easily reach a factor of 3. 
%As a comparison, the ``inner'' sample of Ojha et al. 
%(\cite{Ojha07}) is also shown, i.e. sources inside $-0.5 < l < 0.5$. 

 \begin{figure}[h!]  
\centering  
\includegraphics[angle=0,width=7.5cm]{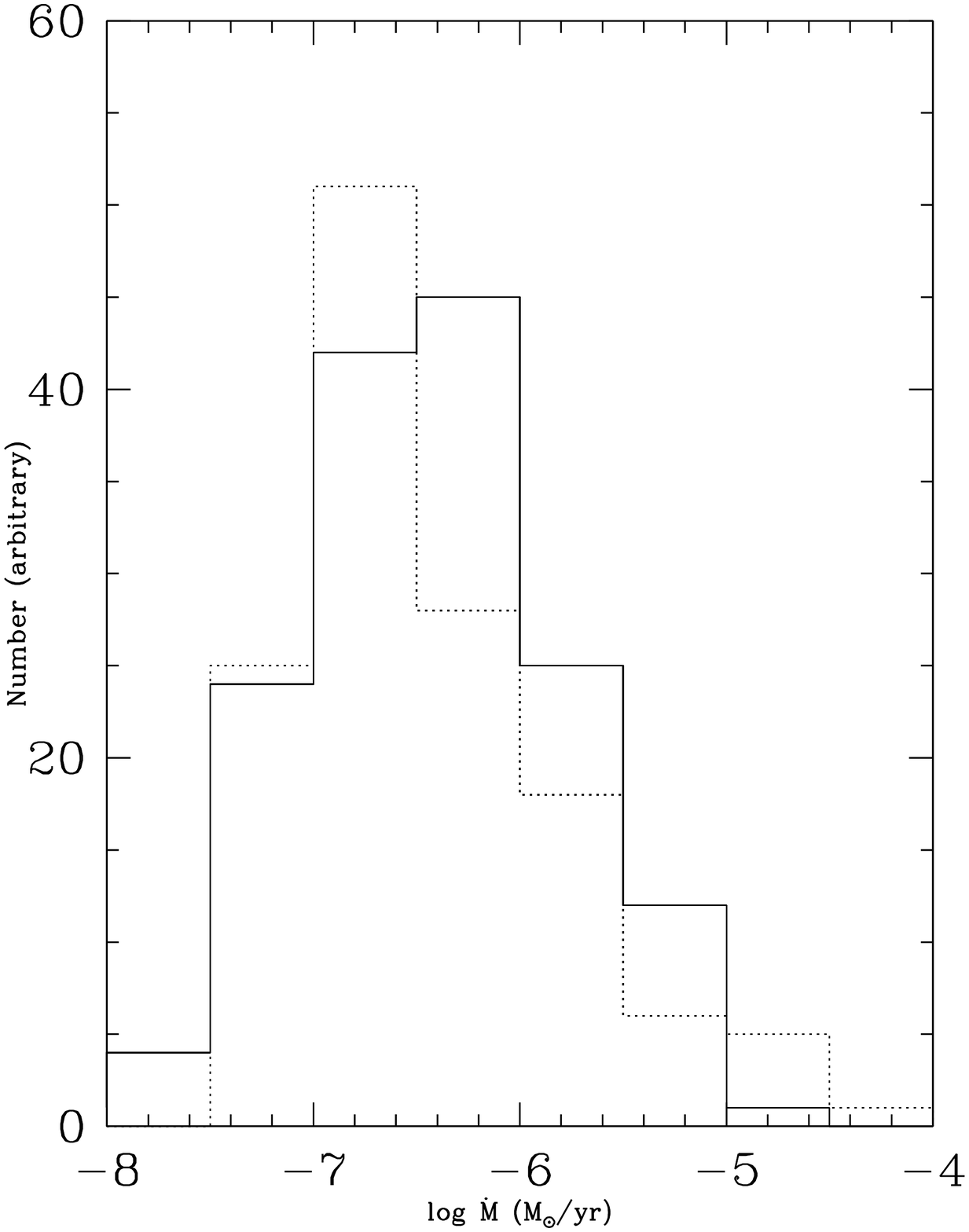}  
%\vspace{3cm}  
 
\caption{Distribution of mass-loss rates for the GC Glass-LPV sample. The mass-loss 
rates are inferred from models (Groenewegen 2006) using the dereddened 
$K-$[15] colour.} 
 
\label{mass-loss}  
\end{figure}  
  
%The derived values of $\dot{M}$ range from $\rm 10^{-8}$ to $\rm 10^{-5}$
%M$_{\odot}$ yr$^{-1}$. The distribution, however, is definitely incomplete
%for $\dot{M}$ $<$ 1 $\times$ $10^{-7}$ M$_{\odot}$ yr$^{-1}$. We want to
%emphasize that the determination of these rates is very uncertain as
%discussed by Schultheis et al. (\cite{Schultheis2003}). The uncertainties
%could easily reach a factor of 3.
  
There is some correlation between mass-loss rates and period, in the sense
that longer periods show higher rates.  Theoretical isochrones by Marigo et
al. (2008) which include mass-loss and pulsation periods predict that there
should be such a correlation; however there is still some disagreement with the
observational data indicating that self-consistent dust models of oxygen-rich AGB stars are in  need of  further improvement.
%Recently, H\"ofner
%2008 presented a new set of hydrodynamical models of M-type AGB stars driven
%by radiation pressure on micron-sized Fe-free silicate grains. These models
%could be used later for comparison with the GALCEN-LPVs. [Not sure if these
%last 2 sentences should be included - it leaves the question as to why we
%have not investigated the fit of these models now. Ian: Well there are very new models
%, I think it will take some ime to get a full grid of these models -letss ay 6months..
%so its just a kind of ``perspective'' -we could leave it out of course.. ]

\section{Luminosities}  
  
We have derived a bolometric magnitude ($M_{\rm bol}$) for each 
Glass-LPV from its de-reddened $K$ magnitude, an assumed bolometric 
correction (BC$_K$), and a distance modulus of 14.5 for the GC.  We
used an approximate relation between BC$_K$ and ($K$--[15])$_{\rm0}$ 
derived by Ojha et al. (\cite{Ojha03}) for AGB stars, using data from Frogel 
\& Whitford (1987).  We refer here to Ojha et al. 
(\cite{Ojha03}) for a more detailed description of the method. These authors conclude that the rms uncertainty in the luminosities of
the Bulge sources that they observed with ISO is about 0.5 mag.
  
\begin{figure}  
\centering  
\includegraphics[angle=0,width=7.0cm]{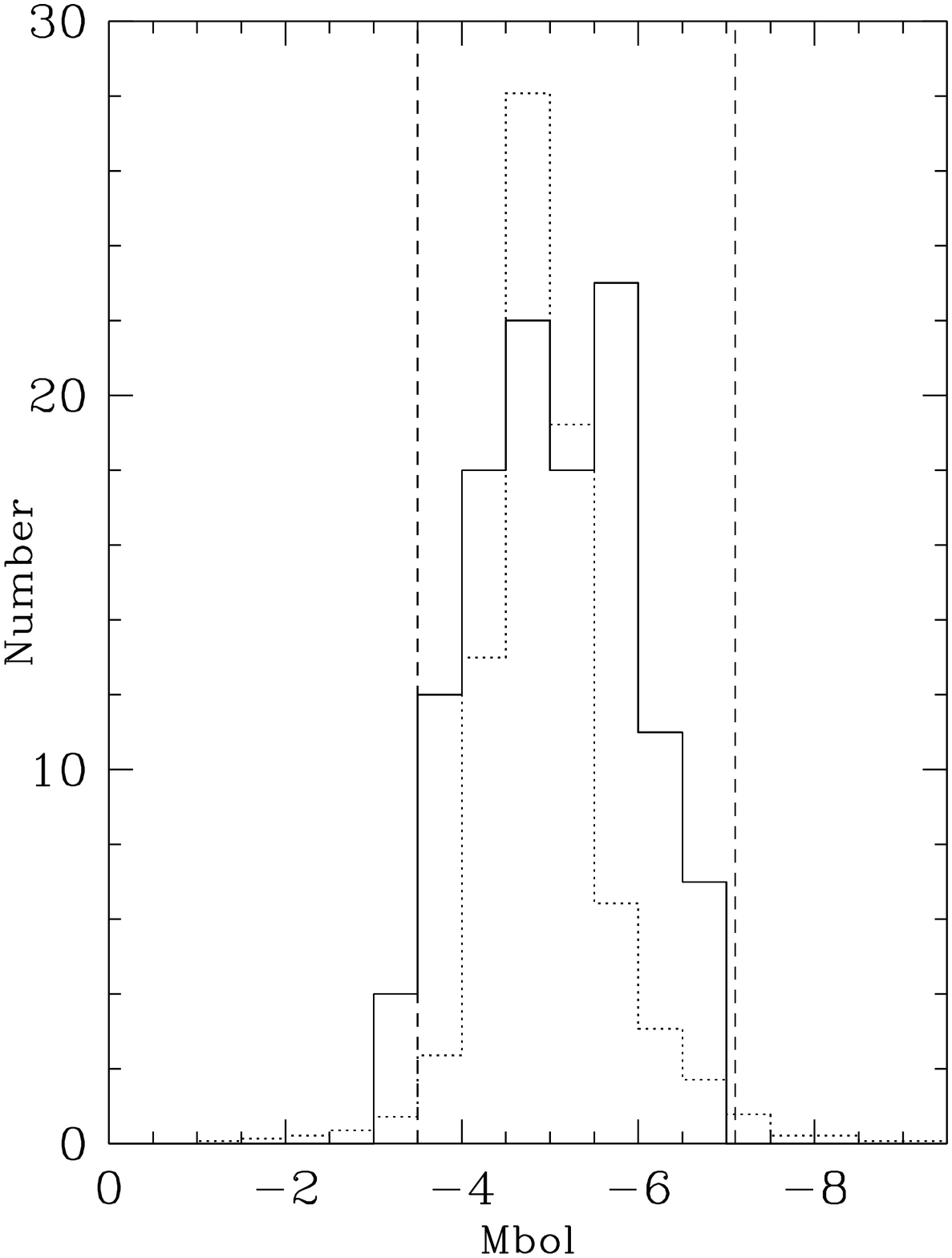}  
%\vspace{3cm}  
 
\caption{Histogram of luminosities of the GC Glass-LPV sample  
(solid line). The luminosities for an inner Bulge sample (Ojha et al. 2007) 
is shown for comparison (dotted lines).  The vertical lines mark the tip of 
the RGB (at fainter $M_{\rm bol}$) and the tip of the AGB (at brighter 
$M_{\rm bol}$). }  
 
\label{histMbol}  
\end{figure}  
  
Fig.~\ref{histMbol} shows the luminosity function of the Glass-LPVs. These  include high luminosity stars, with $M_{\rm bol} < -6$, which are
not observed in the inner Bulge AGB stars (Ojha et al. 2007). However, one should note that the sample
of Ojha et al. (2007) excludes the GC region and are only single-epoch measurements.  The
luminosity distribution of the Glass-LPVs, compared to that of the inner
Bulge AGB stars (Ojha et al. 2007), is also broader and less strongly
peaked. Due to the variability of our sources, the luminosity
distribution will be widened. Thus, compared to the Bulge sample of Ojha et al., we find an excess in the GC of luminous stars.

Such an excess of luminous stars in the GC, compared to Bulge fields such as
Baade's Window, has previously been found in the infrared stellar luminosity
function by Blum et al. (\cite{Blum96}), Narayanan et al. 
(\cite{Narayanan96}), Launhardt et al. (\cite{Launhardt02}), Genzel et al. 
(\cite{Genzel03}), Van Loon et al. (\cite{vanLoon03}), and Figer et al. 
(\cite{Figer04}). In addition, Blommaert et al. (1998) and Wood et al. (1998)
found a similar excess of high luminosity  OH/IR stars in the GC
compared to the Bulge. All these authors conclude that this is generally
evidence for recent star formation in the GC.  Blum et al. (\cite{Blum96}),
Figer et al.  (\cite{Figer04}), and Maness et al. (\cite{Maness07}), among
others, have derived detailed star formation histories for the GC by fitting
models to the observations.  Their studies suggest that star formation has
occurred within the last 10 Myr within the central 30 pc, and that it has
been more or less continuous over the Hubble time.
  
Groenewegen \& Blommaert (\cite{Groenewegen05}) studied the Mira period
distribution of six fields at similar longitudes but spanning latitudes
$-5.8 < b < -1.2 \deg$.  They found for all these fields a similar period
distribution, consistent with the average period in Baade's Window of
$\sim$333 d (Glass et al. \cite{Glass95}).  By comparing their fields with
the LPVs of Glass et al. (\cite{Glass2001}) they found that there is a
significant overabundance of LPVs with periods $\ga$500 d in the GC compared
to Bulge fields at higher latitudes.  This excess of LPVs with long periods
in the GC is not surprising, considering the excess of luminous sources in
the GC generally (see Fig.~\ref{histMbol}), when combined with the
period-luminosity relation.  Groenewegen \& Blommaert (\cite{Groenewegen05})
explain this difference in the period distribution as due to a younger
population in the GC with a higher initial mass in the range of 2.5--3
$M_{\odot }$ compared to 1--2
$M_{\odot }$ for higher latitude Bulge fields.

We conclude  that the GALCEN-LPVs show an excess of luminous, young stars
compared to the galactic Bulge. It is unfortunate that the isochrones in the
[IRAC] bands are not very sensitive to metallicity and/or age; thus
no estimation of initial masses or metallicities in this wavelength range
can be made.
  
%\section{Near-IR spectra}  
  
%We looked at the IRAC colours of the spectroscopic sample of Schultheis et 
%al. (\cite{Schultheis2003}). They performed near-IR spectroscopic follow-up 
%observations of ISOGAL sources with a mid-infrared excess. Besides a 
%majority of AGB stars they found supergiants, red giants and young stellar 
%objects. 
  
%Figure \ref{NIRspec} shows the  colour-magnitude and colour-colour diagram
%of the sample of Schultheis et al. (2003).  The [3.6]$-$[8.0] vs [8]
%diagrams shows a mass-loss sequence starting with the red giants and ending
%up with luminous OH/IR stars. In the [5.8]$-$[8.0] vs [3.6]$-$[4.5] diagram
%the location of the intermediate mass young stellar objects of Allen et al.
%(2004) is indicated.  It is obvious that simple colour criteria alone are
%not enough to separate young stellar objects from AGB stars. Schuller et al.
%(\cite{Schuller2006}) has shown that spatial extension as well as colour
%criteria allows one to better separate YSOs from late-type evolved stars.

%\begin{figure}  
%\centering  
%\includegraphics[angle=0,width=7.5cm]{NIRspectra_bw.ps}  
%%\vspace{3cm}  
 
%\caption{Colour-colour and colour-magnitude diagram for GC sources with
%near-IR spectra from Schultheis et al. (2003). Blue open circles are young
%stellar objects, red crosses plot red giant stars, green filled triangles
%mark LPVs from Glass et al. (2001), open squares indicate known OH/IR stars,
%and filled black circles represent AGB stars (see Schultheis et al. 2003).
%The box shows the location of intermediate mass YSOs (Allen et al. 2004).  }
 
%\label{NIRspec}  
%\end{figure}  

\section{Conclusions}  
 
We have presented a high resolution interstellar extinction map of the
Galactic Center (GC), derived from the Spitzer IRAC catalogue of GC point
sources (Ram\'{i}rez et al. 2008). It combines observations of the RGB/AGB
population with the newest isochrones (Marigo et al. 2008).  We have
improved on previous near-IR extinction maps by using extinctions derived
from mid-IR colours to resolve all regions
of high $A_{V}$.  The maximum value we found is $A_{V} \simeq 90$. By using
this extinction map, we have studied the Spitzer (IRAC) properties of the
long-period variables (LPVs) of in the GC. They follow well-defined
period-magnitude relations in the IRAC bands and are  similar to those observed in the LMC, lending support to the suggestion that the log $P$ vs. [IRAC] relations are universal, independent of metallicity. Further, mass-loss rates for the Glass-LPVs have been determined by using additional data from ISOGAL.  These
LPVs show some correlation between mass-loss and pulsation period as
predicted by theoretical isochrones (Marigo et al. 2008).
Finally, the long periods of the Glass-LPVs as well  as their luminosity 
function agree with the suggestion of Groenewegen \& Blommaert 
(\cite{Groenewegen05}) that one is dealing with a young and more massive 
population in the GC. 
  
\acknowledgements  
 
We want to thank the referee J. Blommaert for his very fruitful comments. We are thankful to M. Marengo for making the colours of AGB stars in the
IRAC bands available to us, and to the GLIMPSE team for sharing data in
advance of publication.  We would like to thank M. Groenewegen for the
fruitful discussion and his comments on the paper This work is based on
observations made with the Spitzer Space Telescope, which is operated by the
Jet Propulsion Laboratory (JPL), California Institute of Technology under
NASA contract 1407.  K.S., R.S., and S.S. are grateful for financial support
from NASA through an award issued by JPL/Caltech. ISG acknowledges receipt of a travel grant under the CNRS-NRF bilateral agreement. M.S. and L.G. acknowledge support by the University of Padova (Progetto di Ricerca di Ateneo CPDA052212).
SG's visit to the Observatoire de Besancon was supported by an EARA-Marie Curie fellowship.  Work at the Physical Research Laboratory is supported by the Department of Space, Govt of India.

 This research has made
use of the NASA/IPAC Infrared Science Archive, which is operated by JPL, California Institute of Technology, under contract
with the National Aeronautics and Space Administration. This publication
makes use of data products from the Two Micron All Sky Survey, which is a
joint project of the University of Massachusetts and the Infrared Processing
and Analysis Center/California Institute of Technology, funded by the
National Aeronautics and Space Administration and the National Science
Foundation.

\end{document}